\documentclass[published]{JHEP3} 

\JHEP{00(2009)000}

\JHEPspecialurl{http://jhep.sissa.it/JOURNAL/JHEP3.tar.gz}

\usepackage{epsfig,multicol,bbm}

\newcommand\fverb{\setbox\fverbbox=\hbox\bgroup\verb}
\newcommand\fverbdo{\egroup\medskip\noindent%
			\fbox{\unhbox\fverbbox}\ }
\newcommand\fverbit{\egroup\item[\fbox{\unhbox\fverbbox}]}
\newbox\fverbbox

\title{Average luminosity distance in inhomogeneous universes}

\author{Valentin Kostov\\
	University of Chicago, Department of Physics, 5640 S. Ellis Ave., AAC 020, Chicago IL 60637\\
	E-mail: \email{valentin@uchicago.edu}}

\received{} 		
\accepted{}		

\preprint{\hepth{9912999}}	

\abstract{Using numerical ray tracing, the paper studies how the average distance modulus in an inhomogeneous universe differs from its homogeneous counterpart. The averaging is over all directions from a fixed observer not over all possible observers (cosmic), thus is more directly applicable to our observations. In contrast to previous studies, the averaging is exact, non-perturbative, and includes all non-linear effects. The inhomogeneous universes are represented by Swiss-cheese models containing random and simple cubic lattices of mass-compensated voids. The Earth observer is in the homogeneous cheese which has an Einstein - de Sitter metric. For the first time, the averaging is widened to include the supernovas inside the voids by assuming the probability for supernova emission from any comoving volume is proportional to the rest mass in it. Voids aligned along a certain direction give rise to a distance modulus correction which increases with redshift and is caused by cumulative gravitational lensing. That correction is present even for small voids and depends on their density contrast, not on their radius. Averaging over all directions destroys the cumulative lensing correction even in a non-randomized simple cubic lattice of voids. At low redshifts, the average distance modulus correction does not vanish due to the peculiar velocities, despite the photon flux conservation argument. A formula for the maximal possible average correction as a function of redshift is derived and shown to be in excellent agreement with the numerical results. The formula applies to voids of any size that: (1) have approximately constant densities in their interior and walls; and (2) are not in a deep nonlinear regime. The average correction calculated in random and simple cubic void lattices is severely damped below the predicted maximal one after a single void diameter. That is traced to cancellations between the corrections from the fronts and backs of different voids. The results obtained allow one to readily predict the redshift above which the direction-averaged fluctuation in the Hubble diagram falls below a required precision and suggest a method to extract the background Hubble constant from low redshift data without the need to correct for peculiar velocities.}

\keywords{cosmological simulations, cosmic web, cosmic flows, gravitational lensing}

\begin{document} 


\section{Introduction}

It is fair to say that the standard cosmological $\Lambda$CDM model is facing a phenomenological crisis. The dark matter carrier has been evading direct detection for decades and the origin of dark energy remains a theoretical puzzle. The most natural candidate is the vacuum state energy, but the flat-space Quantum Field Theory is incapable of calculating it, producing an estimate that is $10^{120}$ times as large as the value suggested by the supernova observations. Presumably, that would be rectified in a fully quantized theory of gravitation which unfortunately does not yet exist. Remaining within the standard General Relativity, there are attempts to explain dark energy as an apparent quantity arising from the averaging procedure that maps the real lumpy universe to the idealized homogeneous Friedman-Lemaitre-Robertson-Walker (FLRW) metric. Such approaches separate into three distinct classes. First, one could average the Einstein equations over spatial hypersurfaces and hope the analogue of the Friedman equation contains a significant effective $\Lambda$ term. Such an idea is conceptually problematic since inhomogeneous universes do not naturally pick up preferred spatial slices to average over, unlike the homogeneous FLRW models. It is shown in \cite{wald} that the result of such averaging depends on the arbitrary choice of slicing and has no coordinate independent meaning - even Minkowsky spacetime could produce an apparent acceleration. A cousin to the first approach is averaging over the null hypersurface of the past light cone of the observer, \cite{lightcone, valerioth}. It is not proven that such averages have a physical meaning and are not simply coordinate quantities. In the second approach, the second order perturbations to the Einstein equations are calculated and treated as an effective energy momentum tensor term. Unfortunately, as pointed out in \cite{wald}, the second order perturbation term is not gauge independent and therefore is not suitable to play the role of an energy momentum tensor in the Einstein equation.

The present paper belongs to the third approach which is completely coordinate independent since it calculates relations only between physically observable quantities such as redshift and luminosity distance (or its logarithmic measure - the distance modulus).  Papers of this type, for example \cite{norwegian},  demonstrate  that the distance modulus - redshift Hubble diagram of type Ia supernovas can be reproduced without any dark energy for an observer occupying the center of a giant ($\sim$ Gpc) underdense spherical bubble. That is possible because the central observer measures a local Hubble parameter that is larger than the global one. The corresponding shift on the Hubble diagram between the true background and the one assumed by the observer allows for fitting the supernova data. Bubbles of such a large scale have significant peculiar velocities away from their center. That creates a fine tuning problem - the observer has to be in the vicinity of the bubble center \cite{offcenter} to observe a small CMB dipole equal to the measured $v/c\sim 0.002$ (Local Group velocity $\approx620$ km/s). This violates the Copernicus principle that we do not occupy a special position in the universe. 

Other papers of the same class studied the collective effect of a configuration of many bubbles/voids of smaller size, thus avoiding big peculiar velocities and significant anisotropies in CMB. That scenario can be conveniently simulated in the "Swiss-cheese" toy model \cite{lemaitre, plebanski, bondi} which has the virtue of being analytically solvable. It is constructed by removing spherical regions from a homogeneous FLRW background (the "cheese") and replacing them with inhomogeneous density distributions with the same gravitating mass (mass-compensated voids). It was used in \cite{valerio, valerioth} for an observer looking along the diameters of a string of aligned voids of radius $350$ Mpc. A positive cumulative change in the distance modulus was found which increased with redshift with respect to the homogeneous background. Although the size of the correction was significant at high redshift, it was not large enough to fit the supernova Hubble diagram and it substituted only partially for dark energy. A smaller correction due to aligned voids with a more mundane radius of $23 h^{-1}$ Mpc was calculated in \cite{brouzakis-eqs}. The result obtained in \cite{valerio} was criticized in \cite{vanderveld} where it was demonstrated to stem from the cumulative weak lensing defocusing of the rays passing diametrically through the aligned voids. The same paper showed that the correction to the distance modulus vanishes when averaged over randomized impact parameters of the rays entering the voids. This conclusion is in accord with an old argument \cite{weinberg} based on the gravitational lensing conservation of the total photon flux. The  implicit geometrical assumptions in \cite{weinberg} have been challenged in more recent papers \cite{mustapha, rose} (and references therein) and the present work will demonstrate they are violated close to the observer due to peculiar velocity modifications of the redshift surfaces.

The studies mentioned above are deficient in several areas. Averaging over directions was not performed in \cite{valerio}. The calculations in \cite{vanderveld} were done using weak lensing theory neglecting the time dependence of the void density and possible nonlinear effects which become significant at low redshifts. The averaging assumed that the ray impact parameters had a uniform probability distribution over the cross-section of a void. That is increasingly true at high redshifts but does not hold in the local neighborhood where the rays have to converge on the observer. More importantly, \cite{vanderveld} averaged only over supernovas residing in the homogeneous cheese. The same was done in a study that obtained an analytical estimate of the corrections to the luminosity distance, \cite{brouzakis}. Such a bias is unjustified, first because supernovas in the voids produce much bigger corrections than the ones in the cheese \cite{valerio}, and second, the observed supernovas occur more frequently in denser regions like the void walls than in the cheese.

The goal of the present paper is to address once again the question whether the direction-averaged distance modulus correction from a configuration of voids really vanishes but this time in the context of a calculation that: (1) is exact, non-perturbative, an includes all possible non-linear effects; (2) is armed with a physically sensible averaging procedure free of ad hoc assumptions about impact parameters \cite{vanderveld} or random cancellations of corrections \cite{weinberg}; and (3) does not neglect the supernovas inside voids. 

The final outcome of the calculations cannot be guessed on the grounds of the previous studies because it may depend on non-linear effects, the choice of averaging procedure, and the supernova selection, none of which was taken into account before. The calculations are performed in Swiss-cheese models with mass-compensated voids of two radii having observational support: $30$ and $300$ Mpc. The cheese is spatially-flat matter-only Einstein - de Sitter (EdS). The observer is placed in the cheese since at present there is no indication that we occupy a void. The first Swiss-cheese model is set up in section 2. The observer shoots past-directed light rays in all directions. The light propagation geodesic equations and the luminosity distance tracing along each ray are discussed in section 3. The physically sensible averaging procedure in this paper assumes that the probability for supernova emission from a comoving volume is proportional to the rest mass in it. Averaging the distance modulus for a single void is discussed in section 4 and it carries the essential characteristics of the procedure for many voids. The main outcome of that section is a simple formula to estimate the maximal average corrections to the distance modulus. Section 5 studies the cumulative correction due to gravitational lensing along a string of aligned voids. Numerical results from averaging within random and simple cubic void lattices are presented in section 6. The effect on the distance modulus produced by large voids of radius $300$ Mpc, which leave a measurable imprint on CMB, is evaluated in Section 7. The summary and conclusions section discusses the answer to the main question addressed by the present study and the practical importance of the obtained low-redshift results for future surveys.

The speed of light in this paper is $c=1$, and times and distances are measured in megaparsecs (Mpc), occasionally giving time in megayears (Myr) for convenience. The gravitational constant $G$ is kept explicit in all equations so that the reader can easily substitute with a favorite value. The usual geometrized units, $G=1$, were used in the numerical calculations.

\section{Swiss-cheese model with voids of radius  $r_v=30$ Mpc}

\subsection{Einstein-de Sitter as homogeneous cheese}

Every reasonable model of the universe is matter dominated in the past while the structure is being formed. Adding a cosmological constant has little impact \cite{bolejko} on the development up until recent times when it starts dominating the matter. That motivates choosing the spatially-flat matter-only Einstein-de Sitter (EdS) model for the homogeneous regions of the Swiss-cheese. The matter includes both visible and dark varieties. This is a convenient playground to explore the question whether voids in the matter distribution can mimic the effect of dark energy. The Hubble constant of the model is $H_0 = 70 \, $ km/s/Mpc. The big bang time is conventionally chosen as $t_{bb}=0$, the current age of the model is $t_0=2/(3H_0)=2857$ Mpc (9312 Myr) and the scale factor as a function of cosmic time is $a(t)=(t/t_0)^{2/3}$. The matter density today is $\bar{\rho}(t_0)=3H_0^2/(8\pi G)$ and as a function of time is:
\begin{equation}\label{rhot}
	\bar{\rho}(t)=\frac{\bar{\rho}(t_0)}{a(t)^3} = \frac{1}{6\pi G \, t^2}\,.
\end{equation}

\subsection{Lemaitre model of a spherical void}

The voids in the Swiss-cheese are modeled with the Lemaitre metric \cite{lemaitre, plebanski} also known as "Lemaitre-Tolman-Bondi solution". It describes a spherically symmetric spacetime filled with an irrotational pressureless ideal fluid (matter, dust). The matter particles are in a free fall under their own gravity tracing geodesics. The zero rotation of the geodesic congruence means the geodesics are hypersurface-orthogonal and the corresponding family of spatial hypersurfaces define a convenient foliation and coordinate system on the spacetime. The resulting coordinates $x^\mu=(t,\, r,\,\theta,\,\phi)$ are matter-comoving and synchronous. The matter energy-momentum tensor in these coordinates is diagonal:  $T_{\mu\nu}=diag(\rho(r,t),\,0,\,0,\,0)$. The metric is given by:
\begin{equation} \label{Lmetric}
	ds^2=-dt^2 + \frac{R'\, ^2}{1+2 E(r)} dr^2 + R(r,t)^2 (d\theta^2 + \sin^2\theta\, d\phi^2)\,,
\end{equation}
where prime denotes a derivative with respect to the radial coordinate $r$. The arbitrary integration function $E(r)$ results from integrating the $G_{tr}=0$ Einstein equation. The areal radius $R(r,t)$ determines the area of a sphere of radius $r$ and $E(r)$ determines the local 3-curvature of the spatial slices. The time coordinate $t$ measures the proper time of the comoving matter. Integrating once the $G_{r r}=0$ Einstein equation leads to the following evolution equation:
\begin{equation} \label{Requation}
	\dot{R}^2 = 2 E(r) + \frac{2 G\, M(r)}{R(r,t)}\,,
\end{equation}
where dot denotes a derivative with respect to $t$ and $M(r)$ is another arbitrary integration function. The above equation alludes to conservation of kinetic plus gravitational energy in Newtonian mechanics where the "energy function" $E(r)$ plays the role of total energy per unit mass. The rest of the Einstein equations connect the "mass function" $M(r)$ to the comoving matter density $\rho$:

\begin{equation}\label{Mequation}
	\rho(r,t)= \frac{M'(r)}{4\pi R^2(r,t) R'(r,t)}\, , \qquad M(r) = \int_0^r 4\pi R(\tilde{r},t)^2 R'(\tilde{r},t) \rho(\tilde{r},t) d\tilde{r}\,.
\end{equation}
The integral can be evaluated for any $t$ and will produce the same $M(r)$. Note, $M(r)$ differs from the comoving rest mass which has an extra factor of $(1+E(r))^{-1/2}$ in the integral. This is an example of a "relativistic mass defect", \cite{plebanski, bondi}. The parametric solution of (\ref{Requation}) for the $E>0$ case of interest is:
\begin{equation}\label{analyticsol}
	R(r,t) = \frac{GM(r)}{2E(r)}(\cosh\eta -1)\,,\qquad
	t-t_B(r) = \frac{GM(r)}{(2 E(r))^{3/2}}(\sinh \eta -\eta)\,,
\end{equation}
where the "big bang time" $t_B(r)$ is another arbitrary function and $\eta>0$ is a parameter. It is natural for inhomogeneous models to have different places with big bang happening at different times.

The metric (\ref{Lmetric}) contains as a subclass all FLRW models which are obtained by setting $R=a(t)\, r\,$ and $E=-kr^2$, $k=const$. In particular, the cheese regions of the present model are spatially-flat and have $E=0$. The voids have to be matched to the homogeneous EdS without tearing of the metric. The appropriate junction conditions \cite{vanderveld} are:
\begin{equation}\label{junction}
	E(r_v)=0\,,\qquad t_B(r_v)=0\,,\qquad M(r_v)=M_{EdS}(r_v)\,,
\end{equation}
where $r_v$ is the radius at which the void merges with a homogeneous region. The first two conditions express an obvious continuation to the EdS values. The third condition means the void is "mass-compensated": the gravitating mass inside it matches the mass in EdS within the same radius $r_v$. The conditions above guarantee that the metric outside radius $r_v$ remains EdS exactly  -  it does not "feel" the presence of the void and the metrics inside different voids do not interfere with each other. Of course, voids in the real world are not restricted within some radius, they interfere and even merge, and their walls accrete mass from outside and regions where the metric stays exactly homogeneous do not exist. The inconvenience is that a real world simulation requires computing of the global metric (or its Newtonian approximation) encompassing the whole spacetime. The mass-compensated voids are a toy-model which is easier to compute because the metrics inside different voids are exact copies of each other. 

The Lemaitre metric is specified by choosing three functions, say $R(r,t_i)$,  $E(r)$, and $M(r)$, where $t_i$ is some initial time and the last two functions determine $\dot{R}(r,t)$ via (\ref{Requation}). The radial coordinate $r$ is just a label for the spherical matter shells and can be chosen at will (gauge freedom). That means the system is specified by only two functions plus a gauge choice for $r$, effected by choosing $R(r,t_i)$. Various possible selections of the two functions are discussed in \cite{hellaby}. 

The $r$ gauge in the present paper is fixed by choosing 
\begin{equation}\label{rgauge}
	R(r,t_0)=r\,.
\end{equation}
It was demonstrated in \cite{acoleyen} that the Lemaitre model can be written from synchronous to perturbed FLRW coordinates (known as "Newtonian gauge"):
\begin{equation}\label{newtonian}
	ds^2= - (1+2\psi)dt^2 + a(t)^2(1-2\psi)(dx^2+dy^2+dz^2)\, ,
\end{equation}
 as long as the peculiar velocities remain small (which is evident later). The void model needs initial conditions in the synchronous coordinates (\ref{Lmetric}) but the astronomical data uses the Newtonian coordinates (\ref{newtonian}). This dichotomy is resolved by the choice in (\ref{rgauge}) which makes the two coordinate systems coincide, approximately,  at time $t_0$ \cite{acoleyen}.  That allows for using Newtonian notions in synchronous coordinates around time $t_0$ and feeding astronomical data of distances, densities and velocities directly to the model. Other models in the literature fix the synchronous $r$ gauge at some initial time much earlier than $t_0$. By the time $t_0$, their synchronous coordinates evolved significantly away from the Newtonian ones which prevents direct Newtonian interpretations and may lead to illusory coordinate effects. For example, light rays that look straight in Newtonian gauge often appear "curved" in such synchronous coordinates.

The present paper uses $\rho(r,t_0)$ and $t_B(r)$ as the two functions specifying the void metric. There is not a standardized definition of what constitutes a void in astronomy.  The so called supervoids are defined as regions free of rich clusters and in the Milky Way proximity have an average radius of about $r_v\sim 46\, h^{-1}$  Mpc $\sim 65$  Mpc (for $H_0=70 \,km/s/Mpc$) \cite{einasto}. Computer void-finding algorithms usually define a void nucleus to be completely free of galaxies leading to breaking down the supervoids into smaller ones of $r_v\sim 20$  Mpc \cite{fairall}. Different algorithms often find voids of different sizes, shapes and numbers in the same region of space \cite{voidcomparison}. A recent paper examined the SDSS 5th release data \cite{visualvoid} using an algorithm that finds voids closest to visual inspection and found an average $r_v\sim36$ Mpc \cite{visualvoid}. It is obvious there is not an universal agreement upon the average void size and the current paper adopts an intermediate value of $r_v=30$ Mpc. The conclusions reached at the end scale with $r_v$ as long as the voids are not significantly nonlinear. The current galaxy density inside voids is very low, below $ 10 \% $ of the average. Density reconstruction of our neighborhood using the peculiar velocity field shows that the total density (dark and visible matter) inside voids drops below $ 0.40\, \bar{\rho}(t_0)$ \cite{dacosta}. The bias between visible and dark matter is still an open question but the present paper will assume that light is a good tracer for matter and set the total matter density inside the void to $ \rho_{in}(t_0) = 0.10\, \bar{\rho}(t_0)$.  Observationally, most of the visible matter is gathered in the void walls and it will be assumed the same applies to the dark matter distribution as well. The wall thickness is set as $\sim 5$ Mpc. The so designed matter density of the void model is
\begin{equation}\label{densityeq}
	\rho(r,t_0)=\bar{\rho}(t_0)
	\cases{
	A_1+A_2\, \tanh[2 (r-25)]-A_3\, \tanh[3 (r-29)] & $,r < 30$ \cr
	1 & $,r \ge 30$}
\end{equation}
and is shown on Fig.\ref{densityF}. The values of the constants to six significant figures are $(A_1, \,A_2, \,A_3)$ $=(0.548006, \,1.25446,\, 0.806455)$. They were determined by requiring density continuity and satisfaction of the mass-compensation condition in (\ref{junction}). The mass function is obtained by evaluating the integral in (\ref{Mequation}) at time $t_0$.

The Lemaitre metric is fixed completely by specifying one more function. This could be the peculiar velocity at time $t_0$, \cite{hellaby}, but it is not clear what a typical velocity profile looks like. The bang time $t_B(r)$ turns out to be a better candidate. After specifying it, one needs to solve for $E(r)$ the two equations in  (\ref{analyticsol}) evaluated at time $t_0$. Eliminating $\eta$ gives
\begin{equation}\label{numericE}
	((1+A_0 x)^2-1)^{1/2} - arccosh(1+A_0 x) = x^{3/2}(t_0-t_B(r))\,,
\end{equation}
where $A_0(r)= R(r,t_0)/(G M)^{1/3}$ and $ x(r)= 2E/(G M)^{2/3}$. The condition for equation (\ref{numericE}) to have a solution $x(r)$ at a given $r$ is derived in \cite{hellaby}:
\begin{equation}\label{solvability}
	t_B(r)>t_{crit}(r)=t_0-\frac{\sqrt{2} \,A_0(r)^{3/2}}{3}\,.
\end{equation} 
Equation (\ref{numericE}) can be solved numerically for $x(r)$ which in turn gives $E(r)$.
All the models with the given density (\ref{densityeq}) are enumerated by the possible functions $t_B(r)>t_{crit}(r)$. The present model sets
\begin{equation}\label{tbeq}
	t_B(r)=0\,,
\end{equation}
which satisfies the junction condition (\ref{junction}) and the solvability condition (\ref{solvability}). 

\FIGURE{\epsfig{file=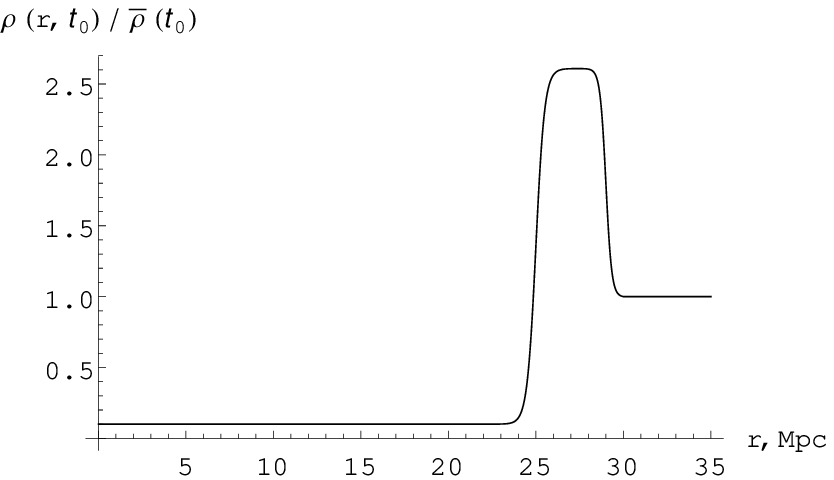, width=7.0cm} \caption{Current void density $\rho(r,t_0)$ divided by the homogeneous density $\bar{\rho}(t_0)$. \label{densityF}} } 

Choosing $dt_B/dr=0$ excites only the growing density perturbation mode \cite{silk, zibin} which is favored by linear perturbation theory in cosmology and increases like $\delta=\delta \rho/\rho$ $\propto t^{2/3} \propto a(t) $ in EdS \cite{peebles}. The decaying mode $\delta \propto H(t)=\dot{a}/a$ \cite{mukhanov} decreases like $\delta \propto t^{-1}$ in EdS. It is neglected in cosmology assuming it had enough time to decay to insignificant levels. Choosing $dt_B/dr\ne 0$ excites a mixture of growing and decaying modes. A common example are Lemaitre models that set the peculiar velocity at initial time $t_i$ to zero which in effect creates a sum of growing and decaying modes with amplitudes $\delta_{grow}(t_i)/\delta_{decay}(t_i)=3/2$ \cite{peebles}. 

The model selected by (\ref{tbeq}) is a pure growing mode. 
Knowing $E(r)$, the metric function $R(r,t)$ can be obtained by either using the analytic solution (\ref{analyticsol}) or by numerically integrating (\ref{Requation}). The second method is advantageous in Mathematica since it produces an interpolation function which is fast to call and whose numeric derivatives are easy to calculate.

\subsection{Void model properties: density, shell crossing, last scattering, peculiar velocity}

The comoving matter density of the void is shown on Fig.\ref{densityratio} and behaves as expected. The ratio of the void wall density to the homogeneous density is increasing at all times.  The wall eventually collapses, reaching infinite density at shell crossing which happens at $t\approx 4500$  Mpc ($14680$  Myr). 
\DOUBLEFIGURE{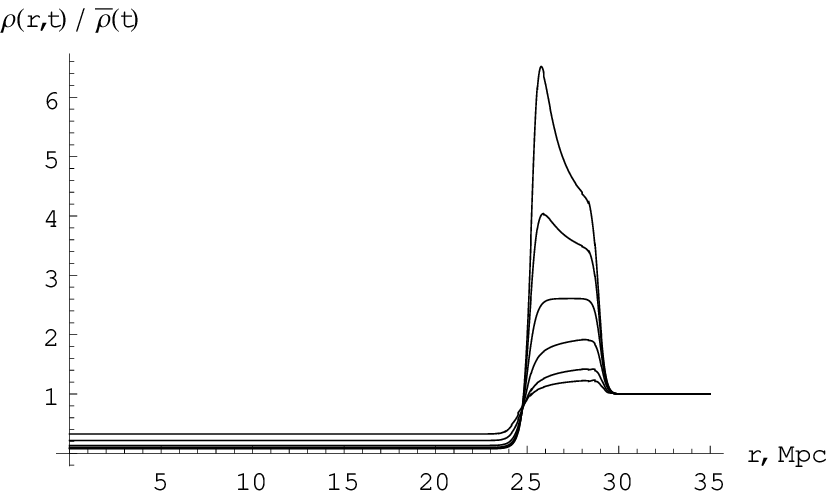, width=7.0cm}{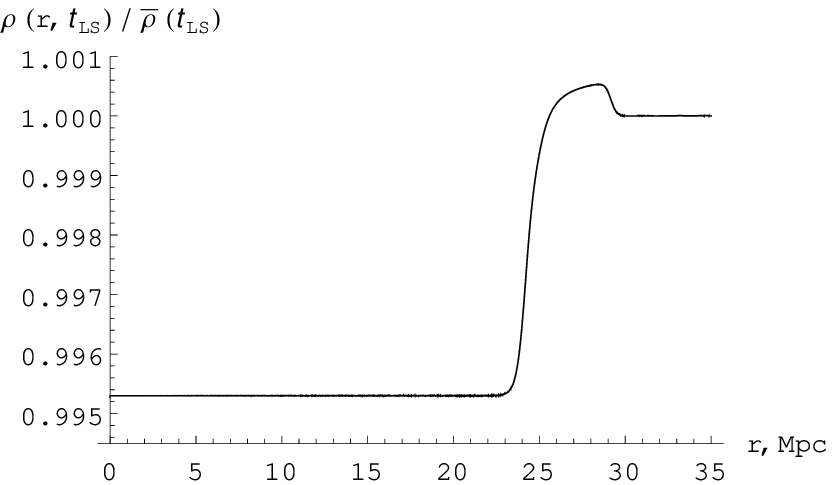, width=7.0cm}{Density ratio $\rho(r,t)/\bar{\rho}(t)$ at times (500, 1000, 2000, $t_0$, 3500, 3900) Mpc (from bottom to top).\label{densityratio}}{Density ratio $\rho/\bar{\rho}$ at time of last scattering.\label{densityls}} 

The time of last scattering in this model, defined as the time at which the EdS scale factor is $a=1/1100$, is $t_{LS}=0.0783 \, Mpc$ (0.255 Myr). The density ratio at last scattering is shown on Fig.\ref{densityls}. The density contrast of the perturbation is of the expected order: $\delta = \rho/\bar{\rho}-1 \sim 10^{-3}$. This $\delta$ applies to the dominating dark matter; the baryonic matter perturbations, suggested by the CMB data, are still $\delta_{bar} \sim 10^{-5}$ at the time of last scattering since the baryons had just decoupled from the photons.  

\FIGURE{\epsfig{file=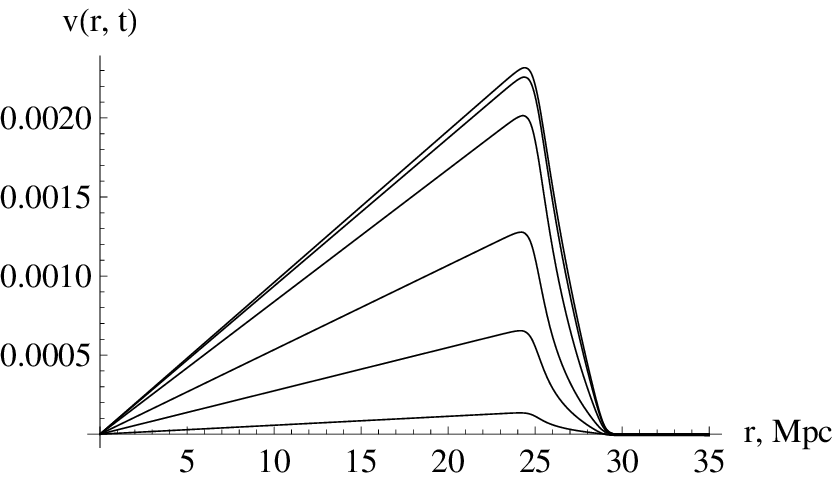, width=7.0cm} \caption{Peculiar velocity profiles at times t=($t_{LS}$, 10, 100, 1000, $t_0$, 4500) Mpc (from bottom to top).\label{v2}}}

The peculiar velocity is defined as the matter velocity with respect to the Newtonian coordinates (\ref{newtonian}). This concept does not apply to synchronous coordinates at all but can be calculated from quantities in them, \cite{acoleyen}:
\begin{equation}\label{velocity}
	v(r,t)=\dot{R}(r,t) - R(r,t) \, \frac{\dot{a}(t)}{a(t)}\,, \qquad v\ll 1 \,,
\end{equation}
Fig.\ref{v2} shows a few velocity profiles at different times.
The peculiar velocity is everywhere positive since in Newtonian coordinates (\ref{newtonian}) the matter underdensity inside the void acts like a repelling gravitational source which accelerates the matter at bigger radius outwards. Even if the initial velocity was negative (towards the center), it will be reversed at later times outwards. For an underdense void interior, a negative initial velocity there is possible only if the model contains a decaying perturbation mode. The current model has a pure growing mode and the peculiar velocity is always positive.

The velocity peak in the profiles can be estimated with standard linear perturbation theory.  The linearized mass continuity equation in Newtonian coordinates gives, \cite{strauss}, $\nabla \cdot v = - a \, \partial \delta / \partial t = - \dot{a}\, \delta \,\partial \ln\delta/ \partial \ln a$. If the void interior is homogeneous, the velocity there is linear (see Fig.\ref{v2}): $v(r,t)=\kappa(t) r$, where $(r,t)$ are Newtonian coordinates (\ref{newtonian}) comoving with the averaged matter, not the synchronous ones. For the growing mode, $\delta \propto a$ in the EdS. Plugging back in the continuity equation obtains $3 \kappa = - \dot{a}\, \delta$ which leads to 
\begin{equation}\label{vlinear}
	v= - \frac{1}{3}\dot{a}\,r\, \delta= - \frac{1}{3}H a\,r\, \delta.
\end{equation}
Fig.\ref{v2} shows that the velocity peaks at synchronous $r\sim 25$ Mpc. Taking that value for the Newtonian $r$ and noting that today $\delta(t_0) = -0.9$ in the void interior one gets $v_{max}(t_0) \sim 1.75 \times 10^{-3}$ when the actual value from Fig.\ref{v2} is $v_{max}(t_0) \sim 2.26 \times 10^{-3}$. The agreement is not bad if one notes that the perturbation is already in a nonlinear regime at time $t_0$. 

The shape of the velocity profiles is explained by the velocity asymptotic behavior close to the big bang when $t-t_B(r) \rightarrow 0$ and $\eta \rightarrow 0$. One can expand the second equation in (\ref{analyticsol}) and then invert the series to obtain 
\begin{equation}
	\eta=6^{1/3} B  - \frac{1}{10} B^3 + \frac{3^{5/3}}{2^{1/3} \,350} B^5 + {\cal O}(B^7)\,, \qquad B(r,t)\equiv \sqrt{2E(r)}\,\left(\frac{t-t_B(r)}{G M(r)}\right)^{1/3}\,.
\end{equation}
Substituting that in the first equation gives
\begin{equation}
	R(r,t)=\frac{G M}{2 E} \left( \frac{3^{2/3}}{2^{1/3}} B^2 + \frac{3^{4/3}}{2^{2/3}\, 10 } B^4 -\frac{27}{1400} B^6+{\cal O}(B^8)\right)\,.
\end{equation}
Inserting the above equation into the peculiar velocity definition (\ref{velocity}) and using $\dot{a}/a = 2/(3t)$ for EdS obtains
\begin{equation}
	v(r,t)= 2 E \left( \frac{t-t_B}{G M} \right)^{1/3} \left(  \frac{(4/3)^{1/3}t_B}{B^2 \, t} + \frac{(3/4)^{1/3}(1+t_B/t)}{5}- \frac{9 (2+t_B/t)}{700} B^2 + {\cal O}(B^4)\right)\,.
\end{equation}
The first term in the above expansion has a time dependence $(t-t_B)^{-1/3} \, t_B/t \sim t^{-4/3}$ for small $t_B\ll t$. One can identify it with the decaying density mode, \cite{peebles}. As mentioned before, that mode vanishes for models with $t_B=const\,$ (=0 by (\ref{junction})). In this case, for times not too far in the future: $t \ll G M /(2E)^{3/2}$  (ensuring $B\ll 1$) , the second term in the above expansion dominates and the velocity profile evolves in time as $\propto t^{1/3}$. This corresponds to the velocity of the growing density mode, \cite{peebles}. The constant shape of the velocity profile is set by $E(r)/(GM(r))^{1/3}$ which is typically triangular since that quantity is zero at $r=0$ and $r=r_v$ and positive in between for $E>0$ models.

\section{Ray tracing}

\subsection{Redshift}
Light rays are null geodesics. The 4-momentum of a photon along such a geodesic inside a void is $K^\mu=\;k_0\;(dt/d\lambda,\, dr/d\lambda, \,d\theta/d\lambda, \, d\phi/d\lambda)$, where $k_0$ is a constant depending on the choice of the affine parameter $\lambda$. The redshift $z$ of a photon emitted by a supernova and observed on Earth satisfies
\begin{equation}\label{prez}
	1+z=\frac{\lambda_{obs}}{\lambda_{em}}=\frac{\omega_{em}}{\omega_{obs}}=\frac{(U\cdot K)_{em}}{(U \cdot K)_{obs}}=\frac{(dt/d\lambda)_{em}}{(dt/d\lambda)_{obs}}\,,
\end{equation}
where $U^\mu$ is the 4-velocity of the emitter and the observer in the comoving coordinates  (\ref{Lmetric}) and dot denotes scalar product. Both the supernova and the Earth observer are assumed comoving with the matter therefore $U^\mu= (1,\,0,\,0,\,0)$ which leads to the last equality. The affine parameter is uniquely defined up to scaling and shifting (affine transformations). One can use those to choose a convenient parameter that satisfies
\begin{equation} \label{Earthobs}
	\lambda_{obs}=0, \qquad (dt/d\lambda)_{obs}=1\,,
\end{equation}
where $obs$ means evaluated at the Earth observer. Substituting that in (\ref{prez}) gives the connection between time and redshift along a light ray:
\begin{equation}
	\frac{dt(\lambda)}{d\lambda}=z(\lambda)+1\,.
\end{equation}

\subsection{Local and global spatial coordinates for ray tracing}
The affine parameter $\lambda$ is chosen to increase with the time $t$ along the ray. The ray's geodesic equations are best integrated by following the ray back in time (decreasing $\lambda$) from the Earth observer ($\lambda=0$) to a supernova ($\lambda<0$). Integration forward in time would require knowing the precise ray direction at the supernova required to hit the Earth and that is hard and impractical to calculate. Note that tracing a ray by decreasing $\lambda$ does not flip the sign of any derivative with respect to $\lambda$!

The fastest method of ray integration inside a void is to use local polar coordinates $(r, \, \theta)$ in the two-dimensional plane containing the ray and the void center, Fig.\ref{diagram}. Due to the symmetric metric (\ref{Lmetric}), a light ray always remains in that plane which can be taken to be $\phi=const$ by properly orienting the spatial coordinate axes. This reduces the number of independent variables by one and simplifies all equations. From void to void, the rays are traced in global spatial coordinates comoving with the homogeneous matter. In the homogeneous cheese, these coordinates are the usual FLRW Cartesian 3-vectors $\vec{x}=(X,Y,Z)$. Inside a void (for $r<r_v$), the global coordinates are Cartesian-like and are connected to the local $(r, \, \theta)$ by the familiar relation:
\begin{equation}\label{globalco}
	\vec{x}=\vec{x}_c +r \cos\theta \,\, \hat{u}  + r \sin\theta\,\, \hat{v}\, ,
\end{equation}
where $\vec{x}_c$ contains the global spatial coordinates of the void center $C$ and the unit 3-vectors $\{\hat{u},\, \hat{v}\}$ form a basis in the two-dimensional plane of the ray, Fig.\ref{diagram}. The above equation does not carry the usual physical sense of conversion from polar to Cartesian coordinates - the spatial metric inside a void is not really Cartesian if written in the global coordinates (\ref{globalco}). Despite that, the coordinate transformation (\ref{globalco}) is mathematically permissible and one is free to use it. 

\FIGURE{\epsfig{file=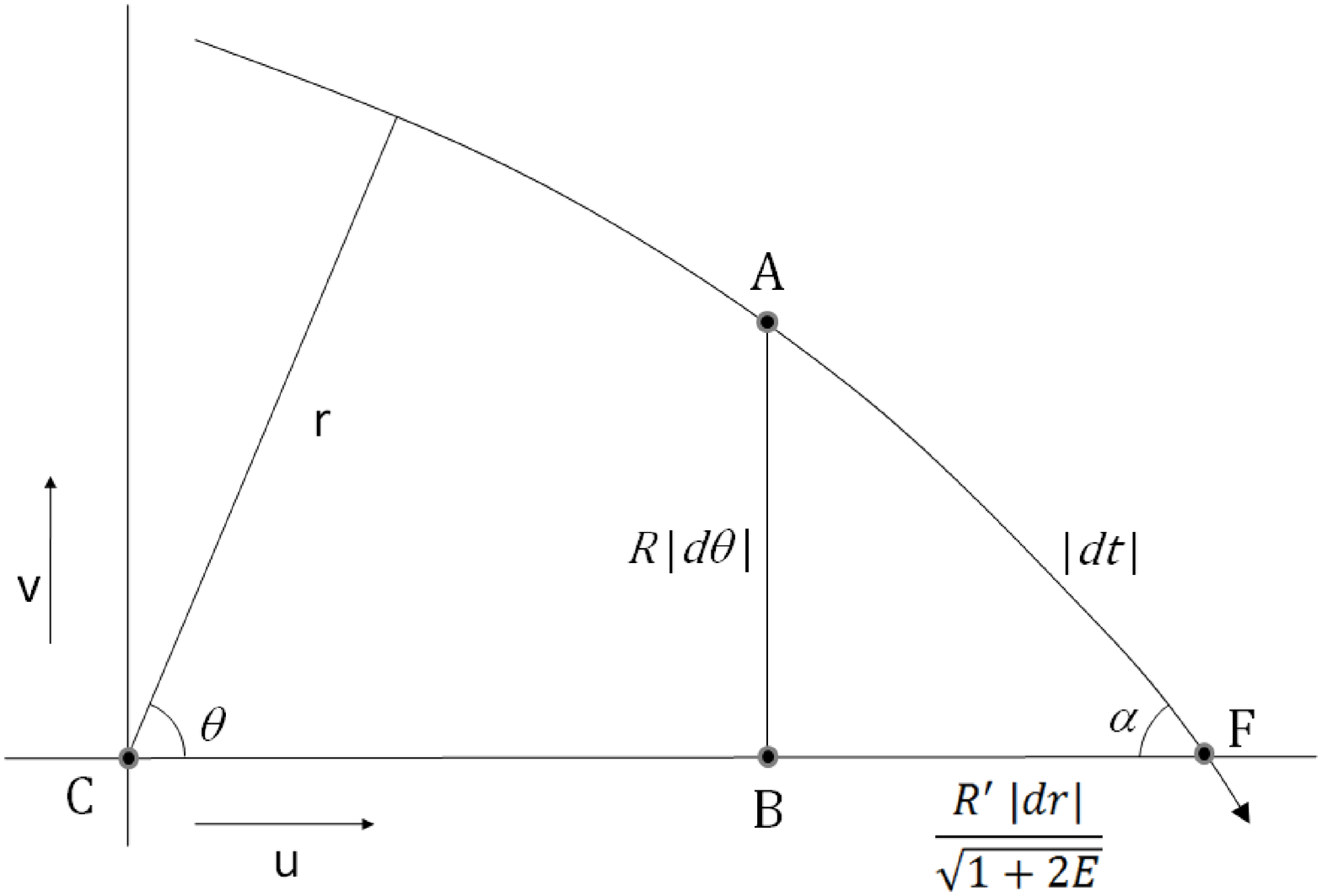, width=9.4cm} \caption{Local coordinates in the ray plane: C - void center, F - point of ray exit, $\{u, v\}$ -  basis unit vectors.}\label{diagram}}

The  $\hat{u}$ vector is chosen to point from the void center to the point $F$ at which the ray  leaves the void (enters the void when propagating back in time). The angle $\theta$ is measured from the u-axis. The  $\hat{v}$ vector is chosen orthogonal to $\hat{u}$ pointing in the half-plane occupied by the ray. The global components of $\{\hat{u},\, \hat{v}\}$, needed to calculate the components of $\vec{x}$, are determined geometrically from the global coordinates of the void center, $\vec{x}_c$, the point $F$, and the ray direction at $F$.

A ray trajectory consists of segments inside the homogeneous cheese and segments inside voids which are glued together in the global coordinates (\ref{globalco}). The final conditions of a cheese-segment determine the initial conditions for integration inside the following void-segment.

\subsection{Void-segments}
The geodesic equations in the plane of the ray ($\phi=const$) are
\begin{eqnarray}
	\frac{d^2 t}{d\lambda^2} + \frac{R' \dot{R}'}{1+2E}\left( \frac{dr}{d\lambda} \right)^2 + R \dot{R} \left(\frac{d\theta}{d\lambda}\right)^2 &=&0\label{eqt}\\
	\frac{d^2 r}{d\lambda^2} + 2 \frac{\dot{R}'}{R'}\frac{dt}{d\lambda}\frac{dr}{d\lambda} + \left(\frac{R''}{R'}-\frac{E'}{1+2E}\right)\left( \frac{dr}{d\lambda} \right)^2 +(1+2E)\frac{R }{R'}  \left(\frac{d\theta}{d\lambda}\right)^2 &=&0\label{eqr}\\
	\frac{d^2 \theta}{d\lambda^2} + 2 \frac{\dot{R}}{R}\frac{dt}{d\lambda}\frac{d\theta}{d\lambda} + 2\frac{R'}{R}\frac{dr}{d\lambda}\frac{d\theta}{d\lambda} = \frac{d}{d\lambda}\left(R^2 \frac{d\theta}{d \lambda}\right)&=&0\,,\label{eqth}
\end{eqnarray}
where $\lambda$ is the affine parameter along the geodesic. The null condition is the first integral
\begin{equation}\label{null}
	\frac{R\,'\,^2}{1+2E}\left( \frac{dr}{d\lambda} \right)^2 = \left( \frac{dt}{d\lambda} \right)^2 - R^2 \left( \frac{d\theta}{d\lambda} \right)^2\,.
\end{equation}
To decrease the computational time, the total differential order of the system can be reduced by using first integrals:
\begin{eqnarray}\label{redeq}
	\frac{dz}{d\lambda}&=&-\frac{\dot{R}\,'}{R\,'}((z+1)^2 - (L/R)^2 ) - \frac{\dot{R}}{R}L \\
	\frac{dt}{d\lambda}&=&z+1\nonumber \\
	\frac{dr}{d\lambda}&=&\pm\frac{\sqrt{1+2E}}{R\,'}\sqrt{(z+1)^2 - (L/R)^2}\nonumber \\
	\frac{d\theta}{d\lambda}&=&-\frac{L}{R^2}\,,\nonumber
\end{eqnarray}
where the last equation is an integration of (\ref{eqth}) and $L$ is the integration constant analogous to angular momentum (naturally conserved in spherical symmetry). The first two equations above are a rewrite of equation (\ref{eqt}) in which $dr/d\lambda$ was expressed from the null condition. The third equation is a square root of the null condition. Its integration has to be stopped at the ray's turning point where $dr/d\lambda=0$ and restarted after that. The sign in the third equation is determined by whether the ray approaches the turning point or goes away from it. 

One needs the initial conditions of integration at point $F$ on Fig.\ref{diagram}. There the ray exits the void forward in time or equivalently enters the void when traced back in time. The triangle $ABF$ consists of physical lengths measured by the matter-comoving observer at point $F$ for an infinitesimal ray segment. The physical length of the ray segment is $dl=|dt|$ because the locally measured speed of light is $dl/|dt|=1$, according to the fundamental axiom of General Relativity. The null condition (\ref{null}) shows that $ABF$ is a right triangle (simply because the coordinates (\ref{Lmetric}) are orthogonal). The angle $\alpha$ is obtained using the end point of the previous cheese segment:
\begin{equation}
	\alpha=\arccos(\vec{dir}\cdot\vec{FC})\,,
\end{equation}
where $\vec{dir}$ shows the spatial direction of the ray {\it traced back in time} and the dot denotes scalar product in the global coordinates (\ref{globalco}). It is easily calculated since the global coordinates of $F$ and $\vec{dir}$ are the end conditions of the previous cheese-segment. One can write the equations
\begin{eqnarray}
dt&=& (z+1) d\lambda \\
\frac{R\,' \,dr}{\sqrt{1+2E}} &=& dt \,\,cos\alpha \\
R \,d\theta &=& dt \,\,sin\alpha\,.
\end{eqnarray}
All the functions must be evaluated at $F$ where $R=r_v a(t)$ and $E(r_v)=0$. After little algebra one gets:
\begin{eqnarray}
z(F)&=&z_F\\
\frac{dt}{d\lambda}(F)&=& (z_F+1)\nonumber \\
\frac{dr}{d\lambda}(F)&=& \frac{(z_F+1)\,\, cos\alpha}{a(t_F)}\nonumber\\
\frac{d\theta}{d\lambda}(F) &=& -\frac{(z_F+1)\,\, sin\alpha}{r_v a(t_F)}\nonumber\\
L=-R^2\frac{d\theta}{d\lambda}(F) &=& r_v a(t_F)(z_F+1)\,\,sin\alpha \nonumber\,,
\end{eqnarray}
where $t_F$ is the time at which the ray hits point $F$. The initial condition used for integration is the one for $L$. 

\subsection{Cheese-segments}
The final point of the previous void segment, corresponding to the smallest $\lambda$ when tracing back in time, is denoted again with $F$. For a cheese-segment, the origin of the local spatial coordinates can be conveniently chosen to coincide with $F$ since there is no center of symmetry analogous to the void center. Rays follow straight lines in the homogeneous cheese. The $u$-axis of the coordinates can be oriented along the ray, $u=\vec{dir}$, which sets $\theta=0$ and $L=0$. The geodesic equations (\ref{redeq}) simplify to
\begin{eqnarray}\label{homeq}
	\frac{dz}{d\lambda}&=&-\frac{2}{3t}(z+1)^2\\
	\frac{dt}{d\lambda}&=&z+1\nonumber \\
	\frac{dx}{d\lambda}&=&-\frac{z+1}{a(t)}\nonumber \,,
\end{eqnarray}
where $a(t)=(t/t_0)^{2/3}$ is the EdS scale factor. The ray trajectory in global coordinates is 
\begin{equation}
	\vec{x}=F + x(\lambda)\,\, \vec{dir}\,.
\end{equation}
The spatial vector $\vec{dir}$ gives the direction along the ray traced {\it back in time}. The minus sign in the third equation above means that $x(\lambda)$ increases when $\lambda$ decreases (going back in time), corresponding to going away from the point $F$ in the direction of $\vec{dir}$ as it should be. The initial conditions are the obvious 
\begin{equation}
	z(F)=z_F, \qquad t(F)=t_F, \qquad x(F)=0\,.
\end{equation}
The corresponding analytic solution is:
\begin{eqnarray} \label{ansolution}
	u&\equiv&1+\frac{5(\lambda-\lambda_F)}{3 \,t_F}(1+z_F)\\
	t(\lambda)&=&t_F\, u^{3/5}\nonumber \\
	z(\lambda)&=&(z_F+1)u^{-2/5}-1\nonumber \\
	x(\lambda)&=&\frac{3\, t_F}{a_F}(1-u^{1/5})\nonumber\,.
\end{eqnarray}

\subsection{Tracing luminosity distance}
The area distance by definition is equal to $\sqrt{A/\Omega_s}\,$, where  $\Omega_s$ is an infinitesimal solid angle subtended by a conical ray at the point source and $A$ is the ray cross-sectional area at some distance from the source. If the ray has an opening angle $\beta\ll 1$ at the source, then  $\Omega_s=2\pi(1-cos\beta)\approx \pi \beta^2$. For rays with a negligible shear, which is the case considered in the present paper, the cross section at some distance is a circle of radius $l$. The angular diameter distance at that position is by definition 
\begin{equation}\label{angdist}
d_A=\frac{l}{\beta} = \sqrt{\frac{\pi l^2}{\pi \beta^2}}=\sqrt{\frac{A}{\Omega_s}}\,,
\end{equation}
equal to the area distance. The reciprocity theorem, first derived in 1933 by Etherington and popularized by Ellis \cite{reciprocity}, connects $r_0$ (original notation) - the area distance for a past directed ray emitted from the Earth observer towards the supernova, to $r_G$ (original notation) - the area distance for a future directed ray emitted from the supernova towards the Earth observer:
\begin{equation}
	r_G=r_0(1+z)\,,
\end{equation}
where $z$ is the redshift between the supernova and the Earth. The luminosity distance to the supernova is \cite{reciprocity}
\begin{equation}\label{lum}
	d_L=r_G(1+z)=r_0(1+z)^2.
\end{equation}
It is quite surprising that these relations are true in arbitrary spacetimes not only in homogeneous ones.

The angular diameter distance along a past directed ray obeys, \cite{brouzakis-eqs}:
\begin{equation}
	\frac{1}{d_A}\frac{d^2 d_A}{d\lambda^2}=\frac{1}{\sqrt{A}}\frac{d^2 \sqrt{A}}{d\lambda^2} = - \frac{1}{2}R_{\mu\nu} k^\mu k^\nu-\sigma^2\,,
\end{equation}
where $\sigma$ is the ray shear, $R_{\mu\nu}$ is the Ricci tensor and $k^\mu=dx^\mu/d\lambda$ is the null tangent vector of the ray. The inverted Einstein equation $R_{\mu\nu} = 8 \pi G (T_{\mu\nu} - g_{\mu\nu} \,T/2)$, where $T_{\mu\nu}$ is the energy-momentum tensor and $T$ is its trace, can be used to calculate the righthand side of the above equation:
\begin{equation}
	R_{\mu\nu}k^\mu k^\nu= 8\pi G(T_{\mu\nu} k^\mu k^\nu - \frac{T}{2} k\cdot k) = 8 \pi G \,T_{00} (k^0)^2\,.
\end{equation}
The last equality follows from $k\cdot k=0$ and the fact $T_{00}=\rho$ is the only non-zero component for a presureless dust in the comoving coordinates (\ref{Lmetric}). Finally, $k^0=dt/d\lambda = z+1$ and the angular diameter equation is
\begin{equation}\label{daeq}
	\frac{1}{d_A}\frac{d^2 d_A}{d\lambda^2}  = - 4\pi G \,\rho(\lambda)\, (\,z(\lambda)\,+1)^2 -\sigma^2\,.
\end{equation}
The initial conditions at the Earth observer are 
\begin{equation}
	d_A(0)=0, \qquad \frac{d \,\,d_A(0)}{d\lambda} = -1\,.
\end{equation}
The last equality is a consequence of the relation $d \,d_A\approx dl$, where the physical distance along the ray close to Earth is $dl=|dt|=|d\lambda|$ due to the local speed of light $c=|dl/dt|=1$ and the normalization choice (\ref{Earthobs}). The negative sign reflects the choice that $\lambda$ decreases tracing the ray back in time, while the distance $d_A$ increases. 

Equation (\ref{daeq}) has to be integrated along with the geodesic equations on each segment of the trajectory. The density $\rho(\lambda)$ is calculated from the first equation in (\ref{Mequation}) in which $r=r(\lambda)$ and $t=t(\lambda)$ are the ones obtained from integrating the geodesic equations. Once the angular diameter distance $d_A=r_0$ is known, the luminosity distance is obtained from equation (\ref{lum}).

The shear is generated by inhomogeneous distribution of matter, \cite{brouzakis-eqs}:
\begin{equation}\label{sheareq}
	\frac{d\sigma}{d\lambda} + 2 \,\theta_e\, \sigma = 4\pi G \rho\left( 1-\frac{3 M}{4 \pi \rho R^3} \right) \frac{L^2}{R^2}\,,
\end{equation}
where $L$ is the angular momentum integration constant in (\ref{redeq}) and $\theta_e=(1/\sqrt{A})d\sqrt{A}/d\lambda = (1/d_A) d\,d_A/d\lambda$ is the expansion of the ray bundle. The righthand side is zero in the homogeneous cheese and the solution is decaying, $\sigma=const \, d_A^{-2}$. In an inhomogeneous void, one can estimate $L^2/R^2=(d\theta/d\lambda)^2 R^2 \sim (\pi/2 r_v)^2 r_v^2 \sim 1$, the bracketed term in (\ref{sheareq}) is $\sim 1$ within the void wall and zero outside and the solution is $\sigma \sim 4 \pi G \rho \Delta \lambda$ with the wall thickness $\Delta\lambda \sim 5$ Mpc. The ratio of the two terms in equation (\ref{daeq}) can be estimated as $\sigma^2/(4\pi G \,\rho) \sim 4\pi G \,\rho \,\Delta \lambda^2$. The current homogeneous density of the model $\bar{\rho}(t_0) \sim 10^{-8}$ for which  $4\pi G \,\rho \,\Delta \lambda^2 \sim 10^{-6}$ and the shear can be safely neglected  in (\ref{daeq}). It plays a significant role only close to mass concentrations like black holes, \cite{brouzakis-eqs}.

\section{Distance modulus averaging for a single void}

Discussing a universe with a single void lays the foundation for understanding the case of many voids. The Earth observer will always be at the origin of the global spatial coordinates. The void center is placed on the $X$ axis, as shown on Fig.\ref{singlebub}.
\DOUBLEFIGURE{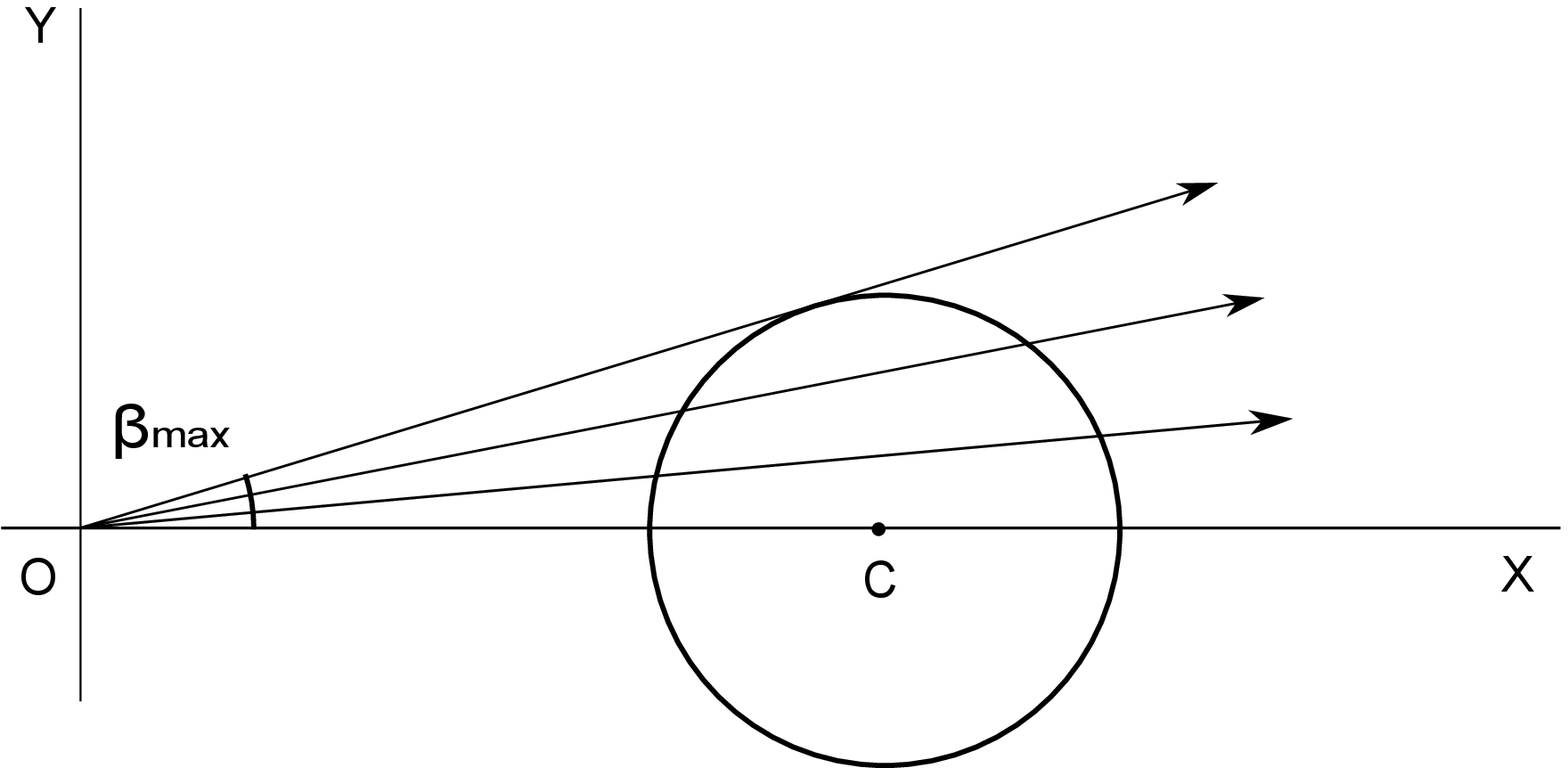, width=7.0cm}{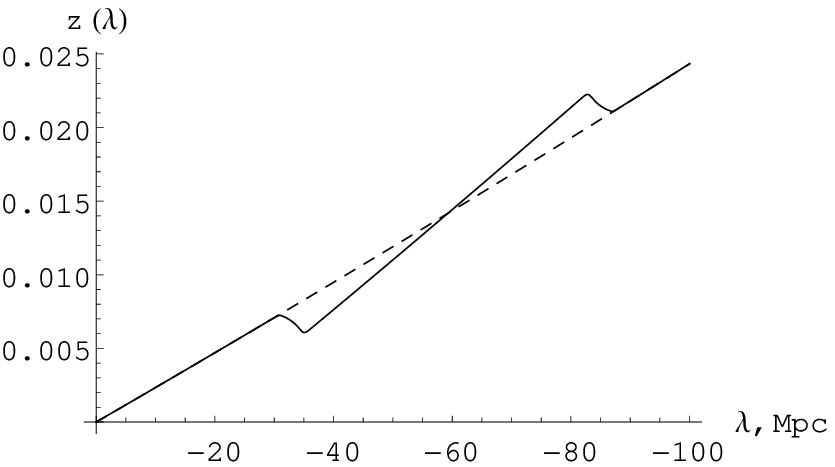, width=7.5cm}{Single void on the X axis penetrated by past-directed rays emanating from the Earth observer at the origin O. The minimal cone for averaging is defined by opening angle $\beta_{max}$.\label{singlebub}}{Solid line - redshift $z(\lambda)$ for a ray emitted at $2\,^\circ$ angle with the $X$ axis and penetrating a void centered at $X=60$ Mpc. Dashed line - redshift $z(\lambda)$ for the homogeneous EdS. \label{zlambda}}
The past-directed light rays look unremarkably straight, at least around time $t_0$. This is due to the chosen spatial coordinates inside the void (\ref{rgauge}), which are approximately Newtonian around time $t_0$. The density inhomogeneities are not sufficiently large to cause a noticeable light deflection in Newtonian coordinates.

\subsection{Redshift surfaces}
  
\FIGURE{\epsfig{file=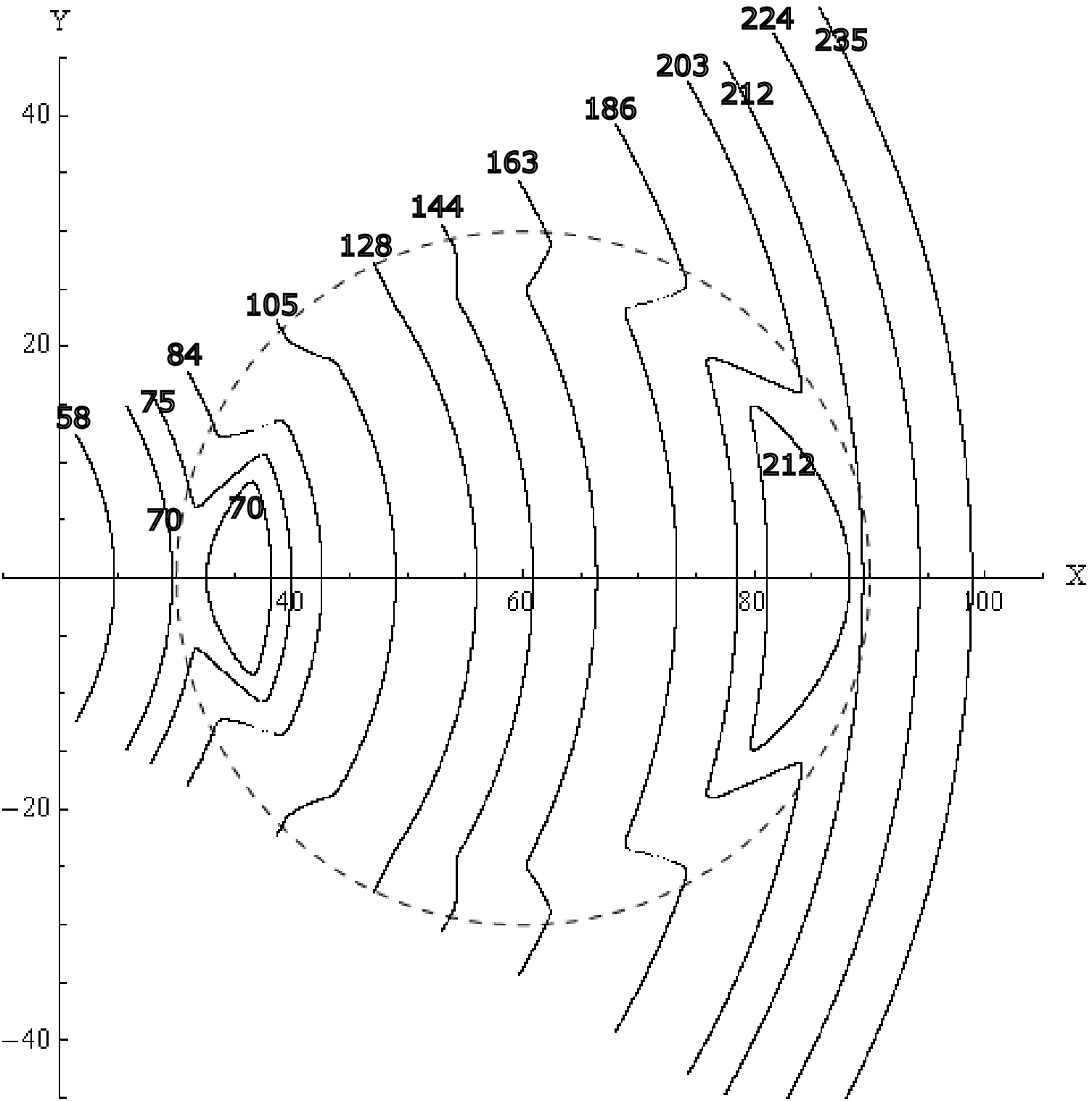, width=7.5cm} \caption{Redshift surfaces for a void centered at 60 Mpc. The bold numbers show $z\,\times\,10^4$. Some surfaces have multiple sheets. The void boundary is shown by a dashed line.}\label{zsurf}}
The computed redshift along a given ray is shown on Fig.\ref{zlambda}. The two "bumps" on the plot are where the ray encounters the front and the back of the void. The peculiar velocity in the front is towards the Earth observer and the redshift is lower; the velocity in the back is pointed away and the redshift is higher. The matter in the EdS cheese has a zero peculiar velocity - it does not feel the presence of the void thanks to the matching conditions (\ref{junction}). As a consequence, the redshift outside the void is indistinguishable from the one of homogeneous EdS as Fig.\ref{zlambda} indicates. The inhomogeneities, through which the ray passes, do cause a minute change in the redshift but it is a second order relativistic effect \cite{brouzakis}. Primary changes in the redshift result from peculiar velocities of the supernova sources inside the void.  

Redshifts in the bumps on Fig.\ref{zlambda} correspond to three $\lambda$ positions along the ray. For a universe filled with many voids, some rays encounter a double bump when they exit a void wall and enter another void nearby. Redshifts in such double-bumps can correspond to five $\lambda$ positions. 

All supernovas with the same redshift lie on a spatial redshift surface. A few of those are shown on Fig.\ref{zsurf} - the actual surfaces are obtained by rotation around the symmetry $X$ axis. The numbers that label the surfaces are the corresponding redshifts multiplied by $10^4$.  Surfaces 70, 75, 203 and 212 consist of three sheets for certain directions - three points giving the same redshift along a ray, corresponding to a "bump" on Fig.\ref{zlambda}. The evolution of the triple surface is traced on Fig.\ref{zsurf}: as the redshift increases to the right, two of the sheets of surface 70 merge on the $X$ axis and open up into surface 75 which gradually straightens out into surface 128 somewhat before the middle of the void. The reverse metamorphosis is seen continuing to the right. Surface 128 folds back into surface 203 which pinches off into the lens-like and open sheets of surface 212. At some distance before and after the void, the redshift surfaces 58 and 235 are just trivial spheres centered on the Earth observer. The surfaces become more symmetric with respect to the middle of the void as its distance to the origin increases.

\subsection{Averaging by solid angle and by mass}
The supernova data displays not the luminosity distance $d_L$ itself but a logarithmic measure of it, the distance modulus: $\mu=5\, \log_{10}(d_L/10 \, pc)\,$. The goal of this paper is to compare the distance moduli of the inhomogeneous universe and the homogeneous background EdS:
\begin{equation}\label{distmod}
	\Delta\mu(z) = \mu(z) - \mu^{EdS}(z) = 5 \, \log_{10} \left[ \frac{d_L(z)}{d_{L}^{EdS}(z)} \right] \, ,
\end{equation}
where $d_L(z)$ is the inhomogeneous luminosity distance (numerically computed by ray tracing) at redshift $z$ and $d_L^{EdS}(z)=3 t_0 ( 1 + z - \sqrt{1 + z})$ is the homogeneous one. For small corrections  $\Delta d_L = d_L - d_L^{EdS} \ll d_L^{EdS}$,  the logarithm can be expanded:
\begin{equation}\label{fraccorr}
	\Delta \mu \approx( 5/\ln10)\, \frac{\Delta d_L}{d_L^{EdS}}\,,
\end{equation}
showing that to first order the change in the distance modulus is given by the fractional change in the luminosity distance.

In an inhomogeneous universe,  $d_L(z)$ depends also on the ray direction and so does $\Delta \mu(z)$ . The correction, averaged over all directions, is obtained by a weighted integration: 
\begin{equation}\label{averagingmu}
	\left<\Delta\mu(z)\right> = \frac{\int \Delta \mu(z; \vec{dir}) dW(\vec{dir})} {\int dW(\vec{dir})}\,,
\end{equation}
where $dW(\vec{dir})$ is the probabilistic weight assigned to the infinitesimal element of the spatial surface or volume over which the averaging is performed. The element is seen in the direction $\vec{dir}$ from Earth. The probability for observing a supernova emission from that element is $dP=dW(\vec{dir}) \,/\int dW(\vec{dir})$.

The previous works assume the probability is proportional to the observational solid angle on Earth, $dW = d\Omega$, and the averaging is over a redshift surface. That is reasonable if all light sources are in the cheese where the density is homogeneous and the probability for a supernova emission is uniform. For practical calculations, the full $4 \pi$ solid angle of the Earth observer is broken down into small solid angles $\Delta\Omega_i$. A ray $i$ with direction inside $\Delta\Omega_i$ is chosen as its representative. The averaging integral is approximated by the sum
\begin{equation}\label{aveang}
	\left<\Delta\mu(z)\right> = \frac{\sum_i \Delta \mu_i \, \, \Delta \Omega_i} {\sum_i \Delta \Omega_i}\,,
\end{equation}
where $\Delta \mu_i$ is the distance modulus correction (with respect to the homogeneous EdS) evaluated at the points where ray $i$ pierces the sheets of the redshift surface $z$. One ray will usually have several such points and a corresponding number of repetitions of $\Delta \Omega_i$. As a result, the total solid angle will be $\sum_i \Delta\Omega_i > 4\pi$ if the sheets are more than one anywhere, which often happens in an inhomogeneous universe. The averaging by solid angle becomes nonphysical if the redshift surface cuts through a void since a significant probability of supernova emission will be assigned to the void interior which is almost empty. The results of this type of averaging will be presented only for reference purposes to show that the nonzero corrections are not artifacts of the more physically appropriate averaging described below. 

The present paper assumes that the probability for observing a supernova emission from a given comoving volume is proportional to the rest mass inside it, $dW=dm$, in agreement with the astronomical views on supernova bias. The rest mass inside the volume specified by an observational solid angle on Earth $d\Omega$, and sandwiched between redshift surfaces $z$ and $z+dz$ is
\begin{equation}
	dm=\sum \rho \,dA\, dl = \sum \frac{\rho\, d_A^2\, (z+1)}{|dz/d\lambda|} \, d\Omega \, dz\, ,
\end{equation}
where the redshift surface area cut off by rays with directions within $d\Omega$ is $dA=d_A^2\, d\Omega$ (from (\ref{angdist})) and the physical distance between the surfaces is $dl=|dt|=(z+1)\, |d\lambda| = (z+1)\, |dz| / \,|dz/d\lambda|$. The sum is over all sheets of the redshift surface, if many. The averaging integral is discretized analogously to the previous averaging procedure:
\begin{equation}\label{avemass}
	\left<\Delta\mu(z)\right> = \frac{\sum_i \Delta\mu_i \, \, \Delta m_i} {\sum_i \Delta m_i}= \frac{\sum_i\, \Delta\mu_i \,\, \rho_i \, d_{A_i}^2\, (z+1) \Delta z \Delta \Omega_i\,/\,|dz/d\lambda|_i} {\sum_i \rho_i \, d_{A_i}^2\, (z+1) \Delta z \Delta \Omega_i\,/\,|dz/d\lambda|_i}\,.
\end{equation}
The sum runs over all rays and all points where ray $i$ pierces the sheets of redshift surface $z$.

\subsection{Maximal average correction in the minimal cone: numerical results}

The real universe does not contain vast homogeneous regions in between the voids. Such regions have $\Delta \mu=0$ and will damp the correction coming from the inhomogeneous void when averaging is performed. The largest possible average $\left<\Delta \mu(z)\right>$ is obtained when the amount of cheese is minimal. With that in mind, the averaging for a single void is restricted to rays inside the minimal cone with an opening angle $\beta_{max}$ on Fig.\ref{singlebub}.  The minimal cone contains the void and a minimal amount of cheese. The obtained average corresponds to the best case scenario when all the voids in that range are exactly at the same radial distance from the Earth observer, forming a spherical shell, and their average corrections add up constructively to the total average.

The void is centered at distance $d=60$ Mpc on the $X$ axis. Due to the axial symmetry, it is sufficient to shoot rays only in the upper half of the $XY$ plane and the full picture is obtained by rotation around the $X$ axis. The angle from zero to $\beta_{max}=\arcsin(r_v/d)=30\,^\circ$ was divided amongst 3000 rays separated by $\Delta\beta=0.01\,^\circ$. Ray $i$ represents the solid angle $\Delta\Omega_i = 2\pi (cos(\beta_i -\Delta\beta) - cos(\beta_i))$ between the cones with opening angles $\beta_i-\Delta\beta$ and $\beta_i$. The redshifts for averaging were 500 values equally spaced on the interval $0.001<z<0.025$. For each redshift, the points where the rays intersect the redshift surface were found and the average correction $\left<\Delta\mu(z)\right>$ was calculated according to (\ref{aveang}) and (\ref{avemass}). The results are shown on Fig.\ref{dmuave}.
\FIGURE{\epsfig{file=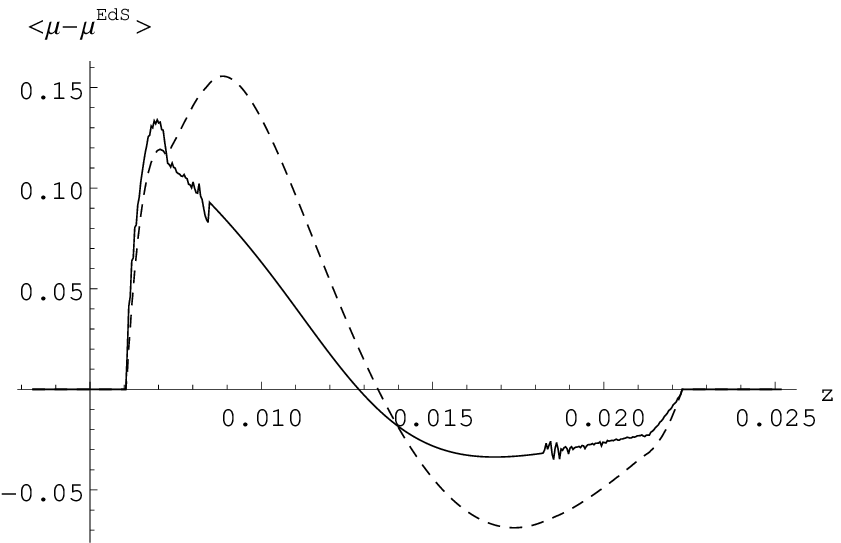, width=7.5cm}\caption{Correction $\left<\Delta\mu(z)\right>$ averaged by solid angle (dashed) and by mass (solid) in the minimal cone of a void centered at 60 Mpc.\label{dmuave}}}
Although the solid angle average is not physically sensible for sources inside the void, it has the same general behavior as  the mass average. This shows the results of averaging are qualitatively independent of the chosen averaging procedure. The redshifts with positive average corrections correspond to surfaces before 128 on Fig.\ref{zsurf}, which are shifted forward with respect to the EdS surfaces, thus having a bigger $d_L$. Surface 70 (redshift 0.0070) is the one with the maximal mass-averaged correction on Fig.\ref{dmuave}. Surfaces after 128 have negative corrections since they are shifted backwards with respect to the EdS redshift spheres.  The smaller magnitude of the negative corrections compared to the positive ones is caused by the bigger amount of cheese in the minimal cone at higher redshifts which damps the average towards zero. The positive and negative corrections become more symmetric with each other for voids at larger distances from the origin - they are completely symmetric in the limit of an infinite distance.

The small spikes appearing on Fig.\ref{dmuave} and later figures are errors from the numerical discretization (\ref{avemass}) of the integrals in (\ref{averagingmu}). Notice they appear at redshifts in the proximity of surfaces $z=0.0084$ and $z=0.0186$ on Fig.\ref{zsurf} which have a segment pointing straight to the observer. It is hard to integrate over such a segment with finite number of rays in (\ref{avemass}). Increasing the number of rays decreases the oscillations but becomes time consuming. The reader should simply ignore the oscillations.
 
\DOUBLEFIGURE{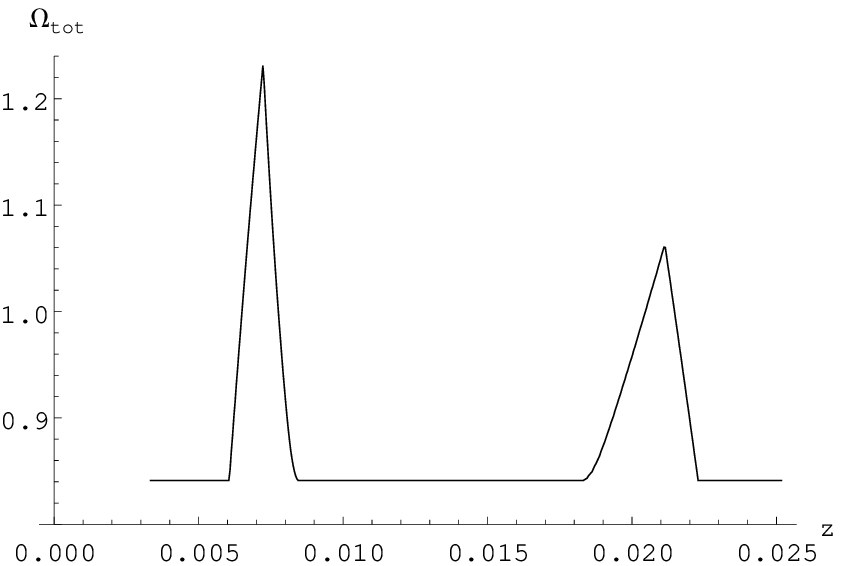, width=7.4cm}{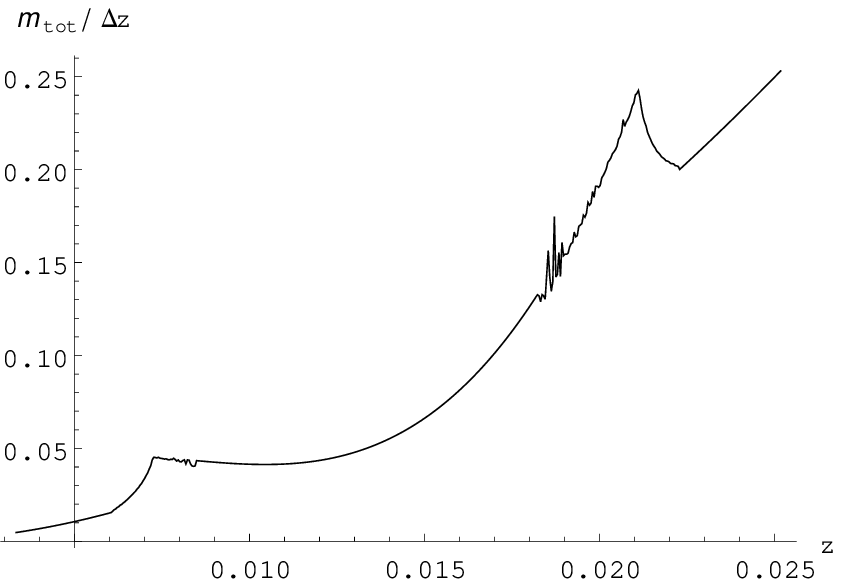, width=7.4cm}{Total solid angle for each redshift surface used for averaging. \label{omegatot}}{Total rest mass for each redshift surface used for averaging.\label{masstot}}

The observer may want to bin the data and find the average correction, $[\Delta\mu]$, over a redshift interval. Naively, one can sum up all the 500 corrections on Fig.\ref{dmuave} and divide by their number to obtain $[\Delta\mu]_{naive}=0.014$ for the angle-averaged and $[\Delta\mu]_{naive}=0.010$ for the mass averaged correction over the whole plotted interval. More appropriately, one has to weigh each $\left<\Delta \mu (z)\right>$ by the total weight of its redshift surface z, shown on Fig.\ref{omegatot} and Fig.\ref{masstot}. For a redshift surface with a single sheet, $\Omega_{tot}=2\pi(1-\cos\beta_{max})=0.84\, srad$, which is the horizontal line on Fig.\ref{omegatot}. The two peaks on that figure correspond to the front and back of the void where redshift surfaces have several sheets leading to overcounting $\Delta\Omega_i$.
The weights on Fig.\ref{masstot} are increasing because there is more mass in the minimal cone at larger redshift. The two peaks in the front and back of the void are again due to overcounting caused by multi-sheet redshift surfaces there. The weighted averages over the redshift interval shown on Fig.\ref{dmuave} are $[\Delta\mu]=0.016$ for angle averaging and $[\Delta\mu]=-0.006$ for mass averaging. The interval averages $[\Delta\mu]$ are significantly smaller in magnitude than the single-redshift averages $\left<\Delta \mu (z)\right>$ on Fig.\ref{dmuave}. That results from the tendency of the positive and negative corrections $\left<\Delta \mu (z)\right>$ from the front and back of the void to cancel out. The cancellation is imperfect close to the Earth observer since the redshift surfaces are not exactly symmetric with respect to the middle of the void. For voids at larger distances, the symmetry improves and so is the cancellation. A perfect symmetry is achieved when the void is at infinite distance from the origin and the rays run parallel to the $X$ axis. Then the interval average would be exactly zero. If one wants to avoid the corrections shown on Fig.\ref{dmuave}, the data must be averaged over intervals bigger than a single void. 

The interval averages calculated above implicitly assume that the supernova detection efficiency is constant throughout the redshift interval. That is usually not true in observational astronomy since supernovas at higher redshift are harder to detect and less in number. The detection efficiency would be especially non constant for big redshift intervals/big voids. That would bias the interval average $[\Delta\mu]$ towards positive values.

\subsection{Maximal average kinematic correction vs. photon number conservation}

The argument \cite{weinberg} based on the gravitational lensing conservation of the total photon flux states that the luminosity distance, averaged over all directions in an inhomogeneous universe, is the same as in the homogeneous counterpart. The inhomogeneities induce focusing/defocusing of light but the argument ascertains these are random along different lines of sight and cancel out in the average. In reality, they are correlated close to the observer and the average does not have to be zero. 

The argument in \cite{weinberg} is implicitly staged in the Newtonian coordinates (\ref{newtonian}). It assumes that the redshift surface in the inhomogeneous universe remains spherical and at the same coordinate distance from Earth as in the homogeneous counterpart. The analysis in the previous sections shows this is not the case because the matter peculiar motion can shift and bend the redshift surface  as shown on Fig.\ref{zsurf}. If the redshift surface shifts along a ray by a comoving distance $\Delta s$ in Newtonian coordinates, the  kinematic correction to the luminosity distance is
\begin{equation}\label{kincorr}
	\Delta d_L \approx (1+z) \Delta s\,,
\end{equation}
There are additional corrections to $d_L$ due to weak lensing or non-linear effects like Rees-Sciama. Neglecting them leads to a very good fit to the numerically calculated maximal average correction at low redshifts which proves the kinematic correction (\ref{kincorr}) dominates there. The other effects become apparent at high redshifts where the kinematic correction has already decayed and for big voids in a deep nonlinear regime or with a decaying mode. Equation (\ref{fraccorr}) shows the distance modulus correction $\Delta\mu$ is proportional to the fractional change in the luminosity distance $\Delta d_L/d_L \approx \Delta s/ s$ ($s$ is the comoving distance to the redshift surface along the ray). It will decrease as   $\Delta \mu \sim 1/s$, because the kinematic shift $\Delta s$ is bounded from above by the void comoving radius.  For distances much bigger than a void radius, the photon conservation argument will apply but only in the approximate sense that with the distance $s$ increasing, the redshift surface is getting closer to spherical and the fractional corrections are getting smaller.

\FIGURE{\epsfig{file=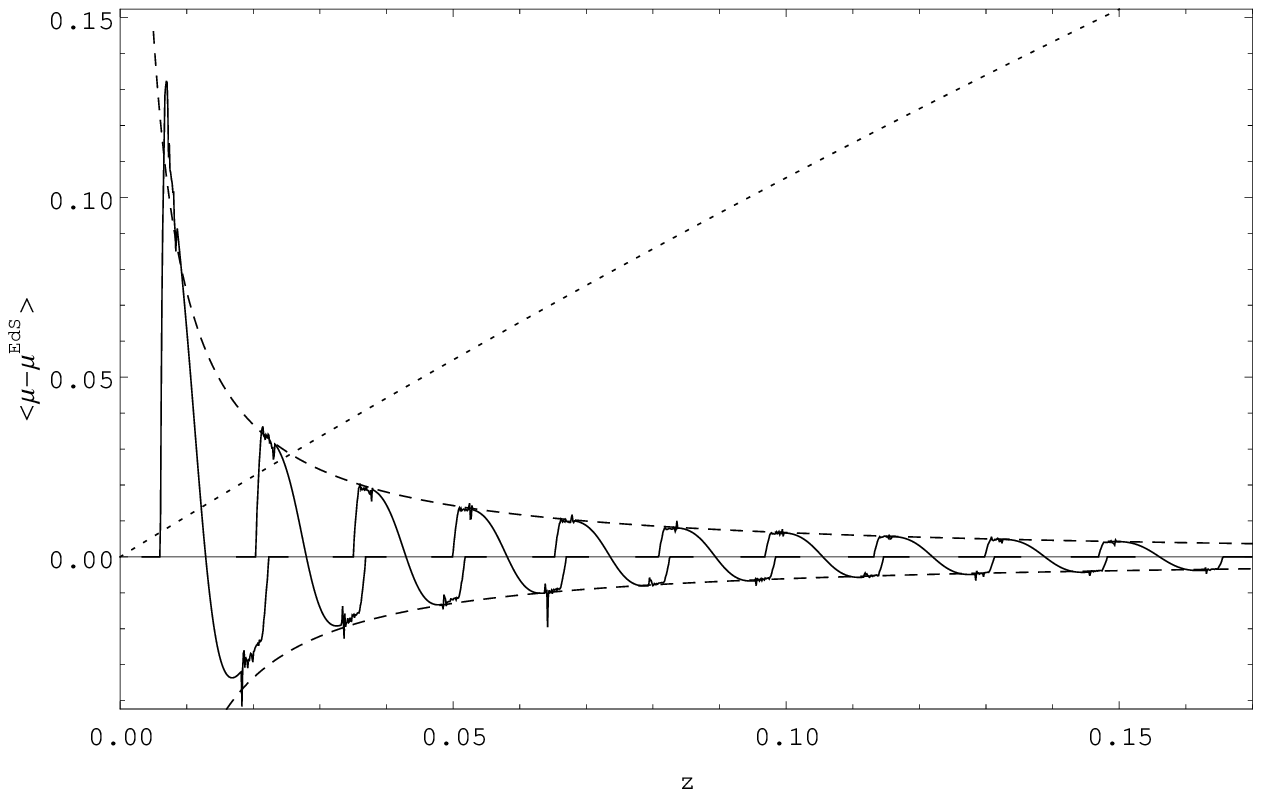, width=15cm} \caption{Solid curves - mass-averaged $\left<\mu-\mu^{EdS}\right>$ in the minimal cone of a void centered at $X=$(60, 120, 180, 240, 300, 360, 420, 480, 540, 600) Mpc. Dashed curves - maximal average correction (\ref{corr}) for $\eta=0.20$ (upper) and $\eta=-0.18$ (lower). Dotted curve shows $\mu^{\Lambda CDM}-\mu^{EdS}$. }\label{maxcorr}} 

From the linear perturbation formula (\ref{vlinear}), the shift of the maximal correction surface $70$ on Fig.\ref{zsurf} is $\Delta s_{max} \propto v_{max}/H \approx a\, r_- |\delta| /3 =r_-|\delta| /3(1+z)$, where $r_-=25\, $Mpc is the radius of the underdense void interior. For models with a linear growing perturbation in an EdS background $|\delta|=|\delta_0|\,a$, where $|\delta_0|=0.9$ is the underdensity in the void interior at present time. The change in the luminosity distance is $\Delta d_L \approx (1+z) \Delta s_{max}  $ $\propto r_- |\delta_0|/3(1+z)$. The corresponding maximal average correction for a given redshift is obtained from (\ref{distmod}):
\begin{equation}\label{corr}
	\left<\Delta \mu(z)\right>_{max} = 5 \, \log_{10} \left[ 1+\frac{\eta\, |\delta_0| \, r_-}{3(1+z)d_{L}^{EdS}(z)} \right] \,,
\end{equation}
where $d_L^{EdS}(z)=3\, t_0 ( 1 + z - \sqrt{1 + z})$ is the homogeneous luminosity distance. The parameter $\eta$ represents the ratio of the maximal average correction to the naive maximal correction corresponding to the maximal peculiar velocity. The value of $\eta$ is not derivable from linear perturbation theory but requires a numerical calculation. It depends on the details of the model (peculiar velocities) and the specifics of the chosen averaging procedure (mass distribution). It was found numerically that $\eta$ does not depend on redshift for the models with a pure growing mode considered here. Models that contain a mixture of growing and decaying mode have a more complicated dynamics; the  redshift surface shapes have a strong time dependence and $\eta\ne const$. 

For small corrections, the logarithm in (\ref{corr}) can be expanded, showing the correction decays like $r_-/s$ or in other words $\propto 1/N_v$ where $N_v=s/(2r_v)$ is the number of void diameters that equal distance $s$.

Fig.\ref{maxcorr} is a composite plot showing the distance modulus correction averaged by mass in the minimal cone of a void placed at various distances from the Earth observer. The void positions do not intersect each other but the corrections overlap since a void can influence a nearby redshift surface that is outside - surfaces 70 and 212 on Fig.\ref{zsurf} gained extra sheets because of the void. The positive and negative corrections get more symmetric at bigger distances from the observer. The dashed curves on Fig.\ref{maxcorr} show the maximal average correction estimated by  (\ref{corr}). The upper curve, enveloping the maximal positive corrections, corresponds to $\eta=+0.20$ and the lower curve, describing the negative corrections, is for $\eta=-0.18$. At small redshifts, the voids enter nonlinear regime and the correction is higher than the linear estimate (\ref{corr}). The dotted curve on Fig.\ref{maxcorr} corresponds to the standard $\Lambda$CDM with $\Omega_m=0.3$ and $\Omega_\Lambda=0.7$ - obviously the average correction cannot substitute for dark energy at redshifts larger than a void diameter from the observer.  

The simple estimate (\ref{corr}) of the ideal scenario maximal correction does not take into account the presence of an excessive amount of cheese in-between the voids nor the fact that the corrections from randomized voids often add destructively as will be seen later. The first effect is an artifact of the Swiss-cheese model - the real universe does not contain vast homogeneous areas between the voids. Both effects will damp the averaged corrections of the model and they will go to zero with redshift faster than shown on Fig.\ref{maxcorr}.

\section{Cumulative distance modulus correction by aligned voids}

\FIGURE{\epsfig{file=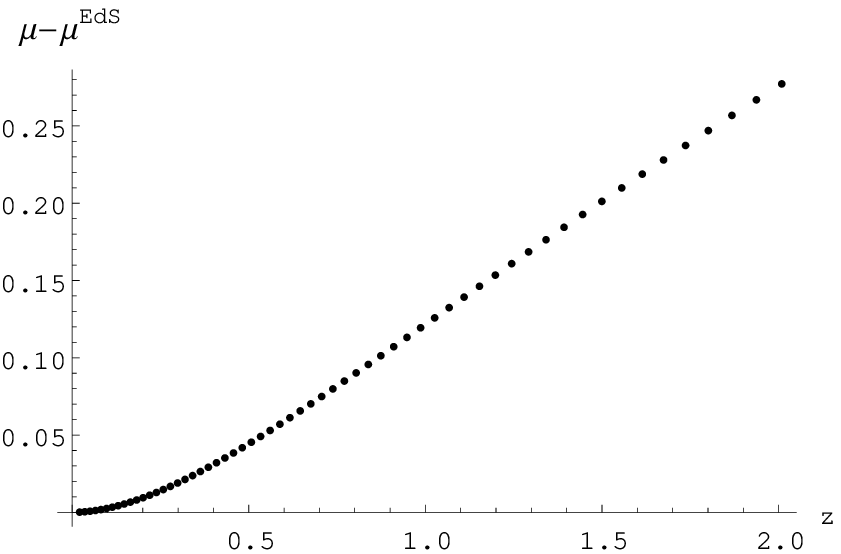, width=6.7cm} \caption{Correction $\Delta\mu$ for a ray at $0.02$ degrees with the $X$ axis penetrating a string of 60 voids centered at $X=2 r_v \, i$ with $i=1...60$.}\label{cumucorr}} 

The cumulative distance modulus correction for a ray at   $0.02$ degrees angle piercing a string of $60$ voids ($r_v=30$) aligned along the $X$ axis is shown on Fig.\ref{cumucorr}. The voids touch but do not intersect. Each dot shows the correction right after the ray exits a void. The correction inside the voids is not shown - it is multi-valued for some redshifts due to multi-sheet redshift surfaces. Including the ray shear in the distance tracing equation (\ref{daeq}) does not influence the result. A correction of a similar magnitude at large redshifts was obtained in \cite{valerio} but for much larger and significantly nonlinear (density contrast in the void wall $\delta(t_0)= 28$ \cite{vanderveld}) voids of radius $350$ Mpc. The effect was ascribed to the cumulative gravitational lensing from the underdense void interiors \cite{vanderveld} and shown to go to zero when averaged over randomized impact parameters. Fig.\ref{cumucorr} demonstrates that a large correction is possible for small voids too. The cumulative correction at a given redshift does not depend strongly on the size or the nonlinearity of the voids but only on how empty they are along the ray. 

One of the main questions addressed in this paper is whether the significant cumulative correction at big redshifts persists in the distance modulus when it is averaged over all directions in a lattice of voids. The answer is no, as shown in the next section, and is obtained without any approximations or assumptions unlike the previous studies \cite{vanderveld}. The more surprising fact is that randomization of the voids is not necessary - even regular void lattices like the simple cubic have a vanishing average correction at high redshifts. Of course regular lattices will still show large cumulative corrections in certain directions, as \cite{valkenburg} demonstrates, but the required void alignment is extremely improbable in the real universe.

\section{Averaging in lattices of $r_v=30$ Mpc voids}

The real universe is permeated by a lattice of voids. Its effect on the averaged distance modulus is calculated by shooting past-directed rays from the Earth observer in directions covering the full solid angle. The luminosity distances at selected redshifts are averaged over all directions via (\ref{avemass}).  

A lattice of 1691 voids was obtained by randomly generating void center coordinates  $(x, \,y, \,z)$ on the interval $-390< x,y,z <390$ Mpc. The voids can touch but not intersect each other. They are not allowed to contain the Earth observer or get closer than $1.5 \, r_v$ to her since that would lead to too big $\left<\Delta\mu\right>$ corrections. The closest void center is at distance $d= 46.9$ Mpc from Earth. The amount of unwanted cheese between the voids depends on the packing efficiency of the lattice, defined as the total volume fraction that is inside voids. The maximal packing efficiency of identical spheres is $\pi/\sqrt{18} \approx 0.74$ achieved in the face-centered cubic and the hexagonal close-packed lattices. The generated random lattice has a packing efficiency of $0.35$ implying too much cheese. Unfortunately, higher packing cannot be achieved by the simple random process described here. 

For comparison, the distance modulus correction was also averaged in a simple cubic void lattice with a cube cell size of $2 r_v$ and a packing efficiency of  $\pi/6 \approx 0.52$. In this case, one can average only over the first octant, the others being identical, thus reducing the computational work $8$ times. The closest void in that lattice is at distance $d=\sqrt{3} r_v \approx 51.96$ Mpc from the Earth observer.

For the random void lattice, the total observational solid angle of $4\, \pi$ was divided approximately equally between $208024$ directions/rays. The resolution in the observational polar angles was $d\theta=d\phi = 2/257$ i.e. the rays could resolve a detail of a comoving size $2$ Mpc at a comoving distance $257$ Mpc (corresponding to redshift $z=0.063$). The luminosity distance was traced along each ray and its values corresponding to $200$ redshift points equally spaced on the interval $0<z<0.08$ were extracted. Those values were used in (\ref{avemass}) to calculate the average for each redshift point. The resulting mass-averaged correction is shown on Fig.\ref{randvoidsave}. For the simple cubic lattice, the $4\pi/8$ solid angle of the first octant was divided between $83599$ directions/rays. The chosen resolution was $d\theta=d\phi = 1.5/345$ i.e. the rays could resolve a detail of size $1.5$ Mpc at comoving distance $345$ Mpc (corresponding to redshift $z=0.086$). The redshift points were the same as for the random lattice and Fig.\ref{scvoidsave} shows the obtained average correction. The thin solid line shows the standard deviation of the corrections at each redshift $\sigma[\Delta\mu](z)=\sqrt{var[\Delta\mu]}$ with the variance $var[\Delta\mu]=\left<\Delta\mu^2\right>-\left<\Delta\mu\right>^2$. The averages $\left<...\right>$ over directions were calculated with the mass weights $dW=dm$.

\DOUBLEFIGURE{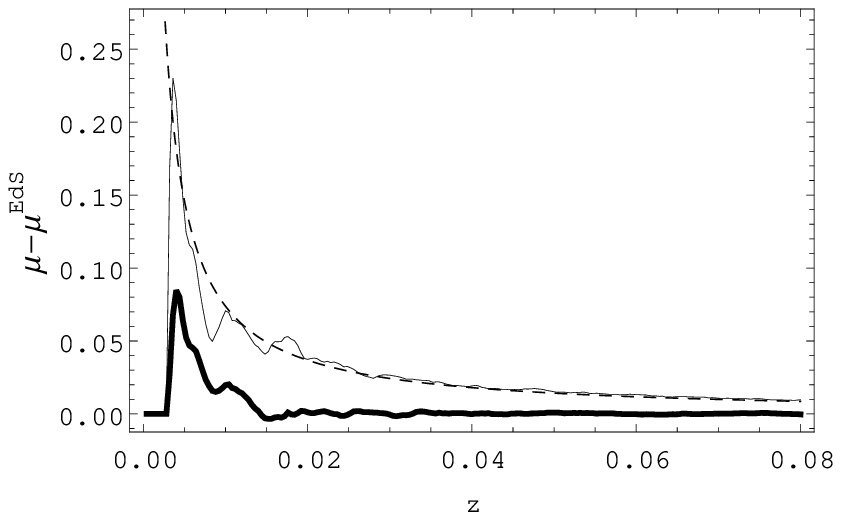, width=7.5cm}{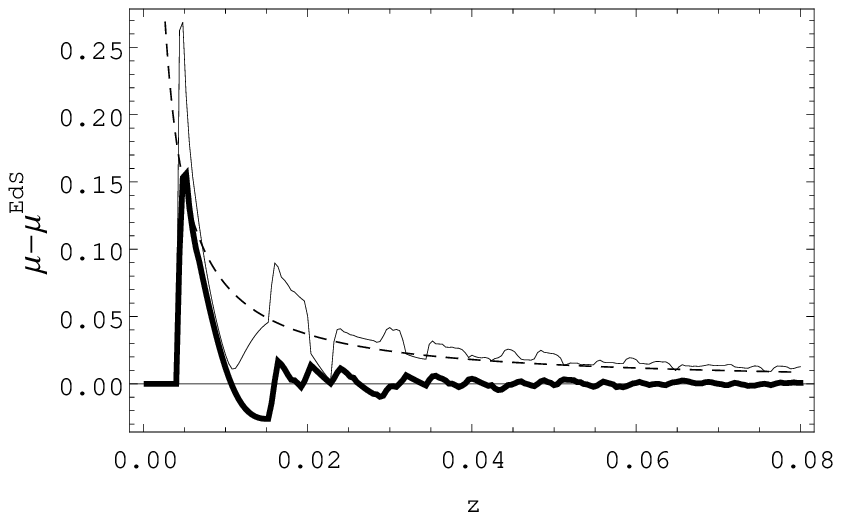, width=7.5cm} {Thick solid curve - mass-averaged $\left<\Delta\mu\right>$ for the random void lattice. Dashed curve - maximal average correction for $\eta=0.20$. Thin solid curve - standard deviation of $\Delta\mu$  at each redshift.\label{randvoidsave}} {Same as the previous figure but for the simple cubic lattice.\label{scvoidsave}}

\DOUBLEFIGURE{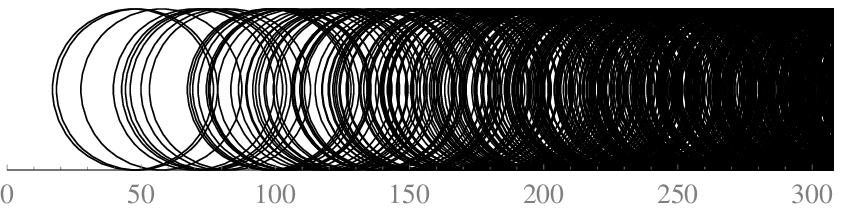, width=7.0cm}{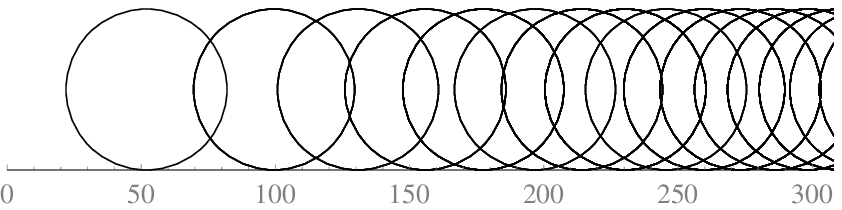, width=7.0cm}
{Distribution of voids in radial distance (Mpc) for the random lattice.\label{randvoids}}
{Same as the previous figure but for the simple cubic lattice. \label{scvoids}}

The average corrections on Fig.\ref{randvoidsave} and Fig.\ref{scvoidsave} are damped below the maximal average correction (\ref{corr}) given by the dashed curve. The first reason for that is that the two lattices have packing efficiencies lower than the one in the minimal cone, which approaches $\sim 2/3$ at distances bigger than several radii. The bigger amount of cheese with zero peculiar velocity in the lattices compared to the minimal cone  damps the average correction towards zero. The much more serious reason for the damping is that whenever a void front and another void back are at the same distance from the Earth, their corrections will tend to cancel out in the total average. The astonishing efficiency of this cancellation damping, especially in the random lattice, at redshifts even as small as a void diameter is puzzling.  This can be understood visually by looking at Fig.\ref{randvoids} and Fig.\ref{scvoids} which show the void distribution in radial distance for the random and simple cubic lattices. Obviously, the front-back cancellation is more severe in the random lattice and at higher redshifts since there is more volume available there and correspondingly more voids per redshift interval. Quite surprisingly, Fig.\ref{scvoidsave} demonstrates that the cancellation occurs very vigorously also in a regular void lattice and even there the cumulative corrections from the previous section do not survive in the directional average. 

The voids in the real universe are not exactly spherical which allows for a tighter packing and more correlated radial void positions close to the observer than in the latices considered here. The expected damping would be between that of the random and the cubic lattice which in this way serve as lower and upper estimators of the average correction. Further discussion of the numerical results is given in the summary and conclusions section.

\section{Averaging in a lattice of $r_v=300$ Mpc voids}

A natural question to ask is whether bigger voids would produce a proportionally bigger average correction than the one obtained so far. 

The current astronomical data suggests that there are structures of size significantly larger than $30$ Mpc. Galaxy count studies \cite{busswell} imply the  Southern local void may have a radius of $200$ Mpc and a density contrast of $-0.25$. The size and underdensity of the largest voids is restricted by the imprint they would leave on the CMB passing through them. A recent paper \cite{granett} found such an imprint by averaging the CMB temperature around voids at a mean redshift $z_v=0.53$. Initially, the signal was attributed to the late-time integrated Sachs-Wolfe effect. Since a significant linear Sachs-Wolfe effect is absent in a flat matter-only universe (see references in \cite{cmbvoid}), the signal was presented as a direct manifestation of dark energy. However, two subsequent papers \cite{cmbvoid, hunt} showed that the measured signal is orders of magnitude too large to be explained with the Sachs-Wolfe effect in $\Lambda$CDM.

The matter density of the voids in this section is chosen to reproduce roughly the average imprint on CMB, shown as a dashed line on Fig.9 in \cite{cmbvoid}. The necessary matter density, shown on Fig.\ref{density300}, is
\begin{equation}
	\rho(r,t_0)=\bar{\rho}(t_0)
	\cases{
	A_1+A_2\, \tanh[0.2 (r-200)]-A_3\, \tanh[0.3 (r-290)] & $,r < 300$ \cr
	1 & $,r \ge 300$}
\end{equation}
with the coefficients  $(A_1, \,A_2, \,A_3)$ $=(0.764715, \,0.349972,\, 0.115257)$, determined by the mass-compensation condition (\ref{junction}) and density continuity. A more precise match to the curve in \cite{cmbvoid} would require a density function with more parameters and is not the goal of the present paper. The bang time chosen for the model is $t_B(r)=0$, implying a pure growing mode. The homogeneous "cheese" between the voids is the same Einstein-de Sitter metric that was used for the small voids. 

\DOUBLEFIGURE{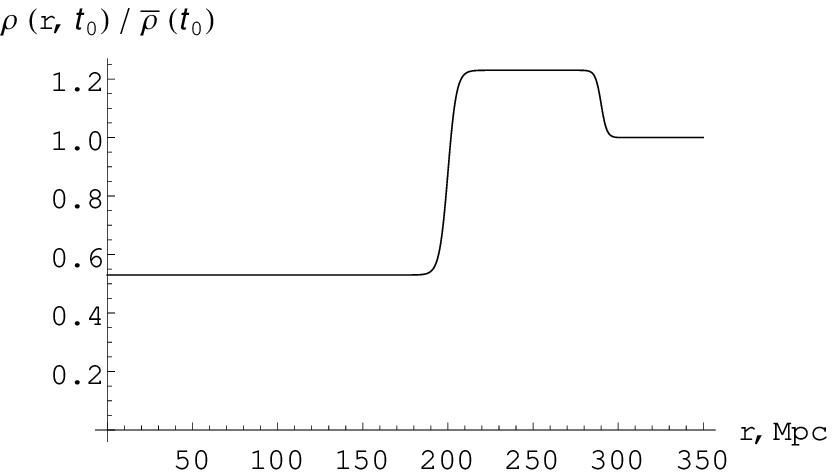, width=7.5cm}{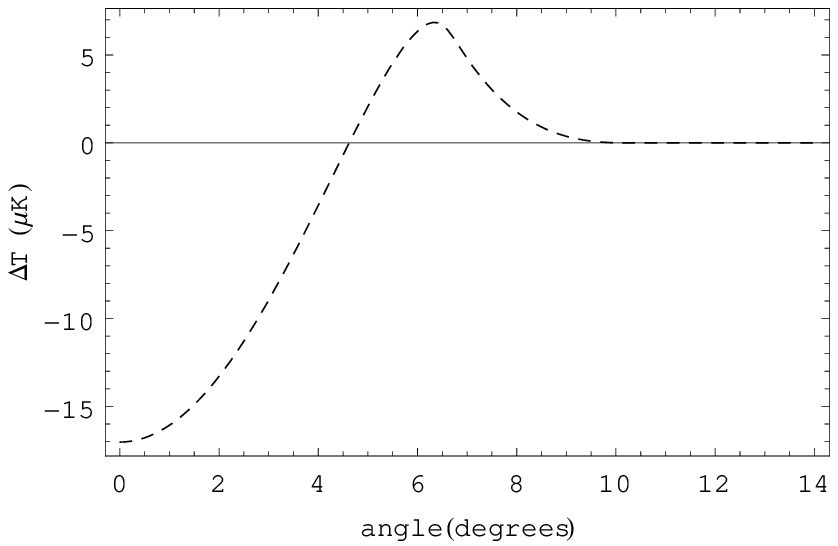, width=7.5cm}
{Density ratio $\rho/\bar{\rho}$ at time $t_0$ for a void of radius $r_v=300$ Mpc. \label{density300}}
{Change in the CMB temperature ($\mu K$) produced by a void of radius $r_v=300$ Mpc centered at redshift $z_v=0.53$.  \label{voideffect}}

Fig.\ref{voideffect} shows the effect of such a void, centered at redshift $z_v=0.53$ ($1633$ Mpc in EdS), on the CMB temperature. It is calculated by shooting rays as shown on Fig.\ref{singlebub} and propagating them to some redshift $z_F$ and time $t_F$ outside the void. The redshift is propagated further to the time of last scattering  $t_{LS}=t_0\, a_{LS}^{3/2}$ using the analytic solution (\ref{ansolution}): $z_{LS}=(z_F+1)(t_F/t_{LS})^{2/3}-1$. The fractional change in the CMB temperature is $\Delta T/T=-\Delta z/z=-(z_{LS}-z_{LS}^0)/z_{LS}^0$, where $z_{LS}^0=1099$ is the unperturbed redshift corresponding to $a_{LS}=1/1100$. 

The effect shown on Fig.\ref{voideffect} and in \cite{cmbvoid} is orders of magnitudes smaller than the one calculated in \cite{valkenburg}. The large $\Delta T/T$ in \cite{valkenburg} is due to: (1) a cumulative effect along special directions from aligned voids in a regular lattice, (2) a significant gravitational blueshift of the rays starting from inside voids at the initial time of the model when the density contrast in the void interior is practically $\delta(t_{init})=-1$, (3) possibly effects produced by the significant non-linearity of the chosen void model \cite{valerio} having a density contrast in the void wall $\delta(t_0)=28$ \cite{vanderveld}. In the real universe, special alignment of more than 2 - 3 voids to produce a cumulative effect on CMB has a vanishing probability. The CMB radiation starts its journey towards us when the dark matter density contrast is of the order $|\delta|\sim10^{-3}$. That produces a gravitational blueshift much smaller than the one exhibited in \cite{valkenburg} . These considerations invalidate the claim in \cite{valkenburg} that voids of radius larger than $35$ Mpc are excluded because they modify significantly the large-scale part of the CMB temperature spectrum. Such voids are actually found \cite{visualvoid} in the SDSS data.

The cumulative correction $\mu-\mu^{EdS}$ along a ray at $0.02$ degrees with the $X$ axis, penetrating a string of $6$ non-intersecting voids centered at $X=r_v \, (2i-1), \,\,i=1...6$ is $\Delta\mu=0.11$ at redshift $z=1.97$. This is about twice as low as the correction for the small voids ($\Delta\mu=0.27$), see Fig.\ref{cumucorr}. The density contrast in the interior of the bigger voids is also about twice as low as for the small $r_v=30$ Mpc voids which indicates that the cumulative effect at a given redshift does not depend on the void size but on how empty the voids are along the ray. Analogously to the case of small voids, the cumulative correction is completely destroyed in the averaging over all directions, even for the regular simple cubic lattice.

\DOUBLEFIGURE{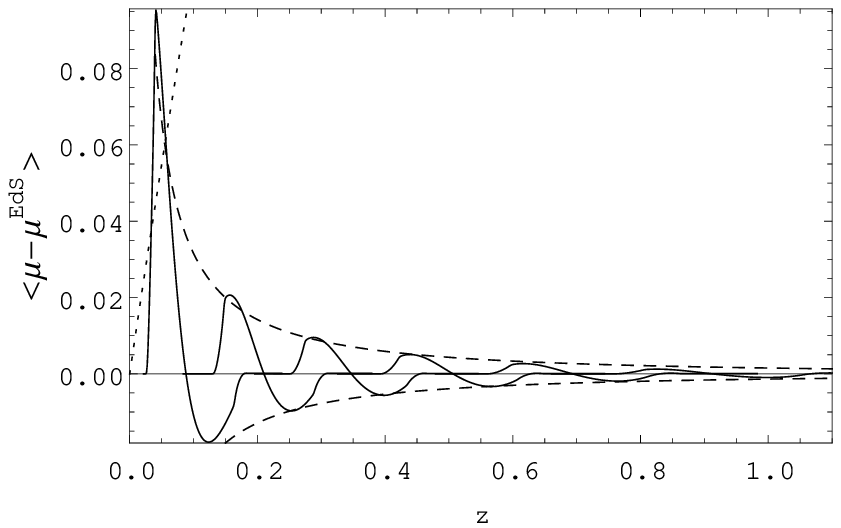, width=7.4cm}{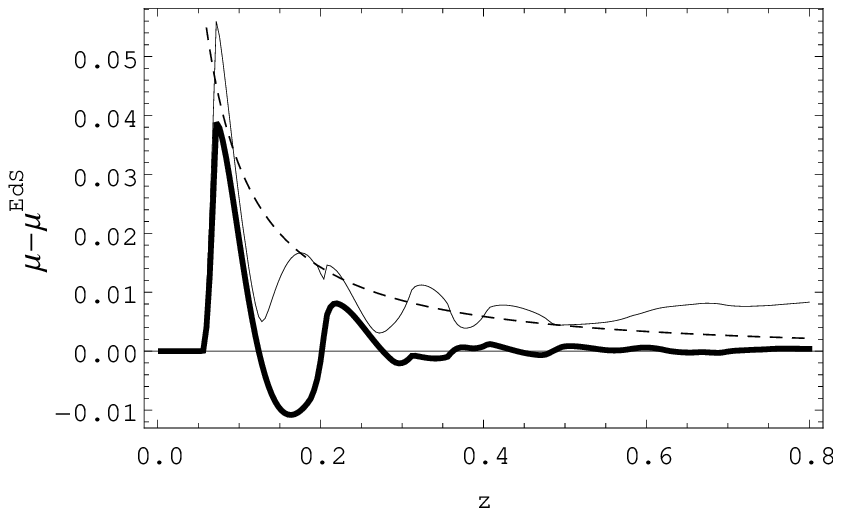, width=7.4cm}
{Solid curves - mass-averaged $\left<\mu-\mu^{EdS}\right>$ in the minimal cone of a void of radius $r_v=300$ Mpc centered at $X=$(400, 800, 1200, 1600, 2000, 2400) Mpc. Dashed curves - maximal average correction for $\eta=0.22$ (upper) and $\eta=-0.20$ (lower). Dotted curve shows $\mu^{\Lambda CDM}-\mu^{EdS}$. \label{maxcorr300}}
{Thick solid curve - mass-averaged $\left<\Delta\mu\right>$ for the simple cubic void lattice. Dashed curve - maximal average correction for $\eta=0.22$ and $r_v=300$ Mpc. Thin solid curve - standard deviation of $\Delta\mu$  at each redshift.\label{scvoidsave300}}

Fig.\ref{maxcorr300} shows the correction averaged in the minimal cone for several positions of a single void. It is counterintuitive that the correction at distance $r_v$ is not bigger than the corresponding one for the small void.  Its magnitude is determined by the ratio $\Delta d_L/r_v \propto  v_{max}/r_-$. The explanation is that although the bigger voids have proportionally larger peculiar velocities, the linear formula (\ref{vlinear}) shows that the ratio $v_{max}/r_-$ is independent of the characteristic void size $r_-$. The dashed curves on Fig.\ref{maxcorr300} are the maximal average correction estimate (\ref{corr}) for $r_-=200$ Mpc and $|\delta_0|=0.47$. The linear perturbation theory estimate (\ref{corr}) envelops the maximal correction solid curves very well except at low redshifts where the voids become mildly nonlinear. It is a little surprising that approximately the same values of $\eta$ from Fig.\ref{maxcorr} work on Fig.\ref{maxcorr300} for an entirely different void model. That suggest some kind of universality of those values for the mass averaging procedure, at least for mass-compensated voids that have a pure growing perturbation mode, are at most mildly nonlinear, and have approximately constant densities in the void interior and walls.

Fig.\ref{scvoidsave300} shows the correction averaged over all directions in the simple cubic lattice. The random lattice was not used since it underestimates the average by having too much cancellation damping as was discussed in the case of small voids. To calculate the average $\left<\Delta \mu\right>$, the simple cubic lattice used for the $r_v=30$ Mpc model was rescaled to the new void radius $r_v=300$ Mpc by multiplying the void center coordinates by $10$. The $4\,\pi/8$ solid angle of the first octant was divided between $51126$ rays with a resolution $d\theta=d\phi = 10/1800$ at distance $1800$ Mpc (redshift $0.60$). The redshift points for averaging were $200$ and equidistant on the interval $0<z<0.8$. The average correction (thick solid line) is damped with respect to the maximal average correction (\ref{corr}) (dashed line) due to cancellation between fronts and backs of different voids, analogously to the case of small voids. Further discussion of the numerical results continues in the summary and conclusions section.

\section{Summary and conclusions}

The paper studied how the distance modulus, averaged over all lines of sight in an inhomogeneous universe, differs from its homogeneous counterpart. The inhomogeneities were represented by random and regular lattices of mass-compensated voids in Swiss-cheese models. Two void layouts were considered with a small ($30$ Mpc) and a big ($300$ Mpc) radius, the first observed in the SDSS data \cite{visualvoid} and the second deduced from its imprint on CMB \cite{granett}.  The Earth observer was put in the cheese but the conclusions will not change significantly for an observer inside one of the voids. For the first time, the averaging of the distance modulus was widened to include the supernovas inside voids. That was made possible by assuming that the probability of supernova emission from a comoving volume is proportional to the rest mass in it. Unlike previous studies \cite{weinberg, vanderveld}, the average correction to the distance modulus was calculated by exact numerical ray tracing without using assumptions about the ray impact parameters or perturbation theory approximations, and the result includes all linear and non-linear effects.

The distance modulus correction, due to gravitational lensing, that accumulates along a ray crossing diametrically a sequence of aligned voids \cite{valerio, brouzakis-eqs, vanderveld} was calculated. The correction increases linearly and becomes quite significant at high redshifts, see Fig.\ref{cumucorr}. It is roughly proportional to the density contrast in the void interior but does not depend on the void radius. It was demonstrated that the distance modulus, averaged over all directions in a lattice of voids, does not show a cumulative correction at high redshifts. That is true not only in a random lattice (see Fig.\ref{randvoidsave})  but also in a regular simple cubic one (see Fig.\ref{scvoidsave} and Fig.\ref{scvoidsave300}) implying that void randomization is not necessary for the destruction of the cumulative correction when averaging, contrary to popular belief. A large cumulative correction is still observed along special directions in a regular void lattice \cite{valkenburg}, but the probability for the required void alignment is vanishing in the real universe.

At low redshifts $z$, perturbation theory predicts that the line-of-sight peculiar velocities $v_r$ are capable of producing significant fluctuations in the luminosity distance $\Delta d_L(z)/d_L \approx - v_r/z$ \cite{hui, haugbolle} ($v_r$ measured in  speed of light units). That leads to a scatter in the local Hubble diagram $\Delta H/ H = -\Delta d_L/d_L$ and, by  (\ref{fraccorr}), to a non-vanishing average correction $\left<\Delta \mu(z)\right>\; \approx \, -2.17 \left<\Delta H / H \right>   \, \approx \, - 2.17 \left<v_r(z)\right>/\,\, z \, $, where the so called peculiar velocity monopole $\left<v_r(z)\right>$ is the line-of-sight peculiar velocity at redshift $z$, averaged over all directions. A measure of the fluctuation magnitude is its average within a sphere of radius $R$ around the observer, $[\Delta H/H]_R$,  which varies with location. Its cosmic variance over all possible observers/locations has been calculated previously in several cosmological models  by N-body simulations \cite{edwin} and by linear perturbation theory using the matter power spectrum \cite{shiturner, shidursi}. Such variances are employed to  predict confidence intervals for the value of $[\Delta H/H]_R$ measured by a random observer. Since the obtained cosmic variance of $[\Delta H/H]_R$ decreases with $R$ \cite{shiturner, shidursi}, with the cosmic mean being zero, the likelihood to measure large average fluctuations decreases with redshift, as expected.  

The cosmic variance of the sphere-average $[\Delta H/H]_R$ over all observers cannot be used to estimate the average over all directions $\left<\Delta H(z)/H\right>$ in the redshift bin $z$, experienced by a \textit{particular} observer.  The results in the present paper allow to calculate $\left<\Delta H(z)/H\right>$ by studying such an observer in a simple model of our local neighborhood. The averages and variances so obtained are not cosmic, thus are more applicable to our astronomical observations. They cannot be calculated from a given matter power spectrum but require a concrete numerical realization of an inhomogeneous universe. To the best knowledge of the author, this is the first time when the non-vanishing direction-averaged correction $\left<\Delta \mu(z)\right>$ is calculated as a function of redshift for universes containing voids. The upper bound (\ref{corr}) for $\left<\Delta \mu(z)\right>$ was motivated using linear perturbation theory and was found to agree with the numerical calculations on Fig.\ref{maxcorr} and Fig.\ref{maxcorr300}.  The parameter $\eta$ depends on the particular form of inhomogeneities and the choice of averaging procedure. The numerically obtained value $\eta \sim 0.20$ indicates that the maximal average correction is $20 \%$ of the naive correction corresponding to the maximal peculiar velocity. This $\eta$ turned out to be a universal constant, approximately independent of the void size, for voids which have approximately constant densities in their interior and walls that are not in a deep nonlinear regime. Due to void randomization, it is expected that the average correction will be suppressed below the maximal $20 \%$ and will tend to zero. The efficiency of that process and at what redshift it sets in has not been studied before. The calculations revealed that the average correction within void lattices is indeed severely damped below the predicted maximal average correction due to cancelations between the fronts and backs of different voids. As figures \ref{randvoidsave}, \ref{scvoidsave}, and \ref{scvoidsave300} demonstrate, the cancelation is surprisingly efficient at low redshifts even in regular lattices and the average correction drops below $0.01$ mag after a single void diameter. Nevertheless, the average correction is not zero close to the observer, indicating that the implicit assumptions of the photon flux conservation argument stated in \cite{weinberg} do not apply at low redshifts. 

The nonzero $\left<\Delta \mu(z)\right>$ is observable, provided there are enough supernovas $N_z$ in redshift bin $z$ to reduce the sampling statistical error of the average: $\sigma[\,\left<\Delta \mu(z)\right>\,] = \sigma[\Delta\mu(z)] /\sqrt{N_z}$, where $\sigma[\Delta\mu(z)]$ is shown as a thin solid curve on figures \ref{randvoidsave}, \ref{scvoidsave}, and \ref{scvoidsave300}. An underdense "Hubble bubble" around us with a Hubble parameter slightly higher than the global one (correspondingly $\left<\Delta \mu(z)\right> < 0$) has been argued for in \cite{zehavi, jha}. However, that claim was not confirmed in \cite{giovanelli, neill} and was shown to depend on the way the supernova magnitudes were extracted from the photometric data \cite{hicken}. Local bubble or not, fluctuations in the distance modulus binned average begin to appear in the newest data sets, Fig.20 in \cite{kessler},  and Fig.7 in \cite{sandage}. 

Significant efforts  in contemporary cosmology are aimed at measuring a  possible time evolution in the dark energy equation of state. That  would require fixing the Hubble constant at low redshifts to a $1 \%$ precision \cite{jha, riess}. Ongoing and future low redshift surveys will boost the number of the observed nearby supernovas sufficiently to bring the sampling error of the average $\sigma[\left<\Delta H / H \right>]\approx \; \sigma[\,\left<\Delta \mu(z)\right>\,] \, / \, 2.17 $ below $1 \%$. In contrast, the directional average itself $\left<\Delta H / H \right> \approx \;- \left<\Delta \mu(z)\right> \, / \, 2.17 $, generated by the coherent peculiar motion, is not influenced by the supernova number and will degrade significantly the errors in measuring the dark energy time dependence \cite{hui, cooray}. Consequently, the peculiar velocities need to be subtracted from the low redshift Hubble diagram.

Several methods are utilized for that.  The Local Group (LG) of galaxies moves as a whole with respect to CMB at a speed $v_{LG}$, the so called peculiar velocity dipole \cite{kocevski, watkins}. The most common way to correct the Hubble diagram for that motion is to adjust the observed redshifts to an observer riding the LG barycenter. The luminosity distance fluctuations seen in that frame are $\Delta d_L/d_L \approx -(v_r-v_{LG})/z$ \cite{hui}  and the velocity dipole will cancel out since the coherent velocity component of a nearby  galaxy with respect to CMB is $v_r^{coh}\approx v_{LG}$. Not correcting the CMB redshifts to the LG frame is permissible for a large supernova sample that covers densely and uniformly all directions: the coherent dipole fluctuation seen in the CMB frame is $\Delta d_L/d_L \approx- v_r/z\approx -v_{LG}\; cos(\theta)\, /z$ and that expression vanishes when averaged over the full solid angle. A further refinement is to correct the LG redshifts for infall velocities in the LG frame caused by the nearby superclusters represented by a crude linear multi-attractor model. That was done in the Hubble Key Project \cite{mould} which measured the Hubble constant to a $9\, \%$ precision, but it was not very efficient at reducing the scatter in the low redshift Hubble diagram as seen on Fig.4 of \cite{finalhubble}. Accordingly, the philosophy of the Hubble Key Project was to extract the Hubble constant mainly from secondary distance indicators at high redshifts. A finer method to correct for peculiar velocities is to calculate them from the observed galaxy distribution utilizing linear perturbation theory and choosing a galaxy-dark matter biasing parameter that minimizes the scatter on the Hubble diagram. That technique does reduce the scatter \cite{radburn} but using it to infer the Hubble parameter from low redshift data produces a controversially high value of $H_0= 85 \,km\, s^{-1}\,Mpc^{-1}$ \cite{willick}. To avoid unreliable velocity corrections, many recent surveys \cite{kessler, riess, sandagehubble} simply chose to include only objects above a certain redshift, $z>z_{min}$,  where the peculiar velocities are considered insignificant compared to the Hubble flow. The Hubble constant and the dark energy equation of state parameter $w$ inferred from the low redshift data are sensitive to the choice of $z_{min}$ \cite{jha, kessler}. As discussed in \cite{kessler}, there is a wide variation in the values selected for $z_{min}$ in the literature, ranging from $0.01$ \cite{sandagehubble} up to $0.023$ \cite{riess}, which indicates that there is not a universally accepted prescription.

Figures \ref{randvoidsave}, \ref{scvoidsave}, and \ref{scvoidsave300}  of the present paper allow to select $z_{min}$ readily. The required $\Delta H/H = 0.01$ precision corresponds to $\Delta \mu_{goal} = 0.0217$. The sampling error of the average is $\sigma[\Delta\mu(z)] /\sqrt{N_z}$, where $\sigma[\Delta\mu(z)]$ is given by the thin solid curves on the plots and $N_z$ is the number of objects/supernovas in redshift bin $z$. Provided there is sufficient statistics, the sampling error will fall below $\Delta \mu_{goal}$. In that case the limiting value $z_{min}$ is determined by the average $\left<\Delta \mu(z)\right>\ $ itself - a natural choice is the redshift at which the maximal average correction (\ref{corr}) (the dashed curves) drops below $\Delta \mu_{goal}$ or the redshift corresponding to a void diameter (where $\left<\Delta \mu(z)\right>\ \sim 0.01 $ due to front-back void cancellation), whichever is lower. In comparison, estimates based on the sphere-average $[\Delta H/H]_R$ are more pessimistic: its cosmic variance  in the $\Lambda$CDM model drops below $1\%$ at $z_{min}=0.05$ \cite{shidursi} (linear estimate). If our neighborhood contains voids of the size observed on the CMB imprint \cite{granett}, the required $z_{min}$ would be much larger, as Fig. \ref{scvoidsave300} indicates.  

Excluding the objects below $z_{min}$ reduces the size of the sample and proportionally increases the statistical error \cite{cooray}. The present paper suggests it is possible to preserve the low redshift data, instead of throwing it away, by collapsing it into an interval average $[\Delta \mu]$ over a redshift interval containing a void diameter. These averages turn out very close to zero: $[\Delta \mu]=0.003$ ($z=0\, \ldots \,0.025$ on Fig.\ref{randvoidsave}), $[\Delta \mu]=0.007$ ($z=0\, \ldots \, 0.025$ on Fig.\ref{scvoidsave}), $[\Delta \mu]=0.002$ ($z=0\, \ldots \,0.25$ on Fig.\ref{scvoidsave300}). That stems from the cancellation between positive and negative lobes in $\left<\Delta \mu(z)\right>$ and the mass averaging giving higher weights to higher redshifts where $\left<\Delta \mu(z)\right>$ is closer to zero. A  vanishing $[\Delta \mu] \approx 0$ means $[\mu]\approx[\mu]^{EdS}$:

\begin{equation} \label{extracthubble}
	\sum_i W_i \left<\mu(z_i)\right> \approx \sum_i W_i \; \mu_i^{EdS}(H_0, z_i),
\end{equation}

where $W_i$ is the mass weight assigned to redshift bin $z_i$, $\left<\mu(z_i)\right>$ is the distance modulus in that bin averaged over all directions, and the sum is over all the bins in the redshift interval. The lefthand side of the formula is calculated from the observational data. The weight $W_i$ is proportional to the rest mass inside the redshift bin which can be estimated as the corresponding mass in the homogeneous background model. Unlike a traditional Hubble diagram that weighs all redshift bins equally, here $W_i \propto z_i^2 \, \Delta z_i$ at low redshifts. The righthand side of (\ref{extracthubble}) depends solely on the Hubble parameter $H_0$ at low redshifts and on additional cosmological parameters of the background model at higher redshifts. It allows one to extract the value of $H_0$ from the low redshift data. This method will become possible for future surveys that: (1) have a dense full-sky coverage to calculate the average correction over all directions $\left<\Delta \mu(z)\right>$; and (2) contain a large number of objects in each redshift bin so that $\left<\Delta \mu(z)\right>$ is readily observable, not swamped by the sampling error.

If the observer is inside a void, the average correction $\left<\Delta \mu(z)\right>$ starts with a negative lobe unlike the so-far considered case of an outside observer. A simple illustration of that is an observer at the center of a void, which will measure $\Delta \mu(z) = \mu(z) - \mu^{EdS}(z)<0$ for supernovas inside the void since the matter there is expanding faster than the background and it takes a smaller distance $d_L$ to achieve the same redshift. The interval average $[\Delta \mu]$ over a void diameter would still be very small since the mass averaging scheme adopted in this paper ascribes lower weights to low redshifts. Additional studies are needed to investigate the behavior of $[\Delta \mu]$ in the presence of superclusters - the other type of major inhomogeneities encountered in the universe. A plausible guess is that $[\Delta \mu]$ vanishes on a redshift interval encompassing a supercluster, justifying the validity of (\ref{extracthubble}) in that case as well.   

\section{Acknowledgements}
The author is indebted to the financial support of the USA Department of Energy.

\end{document}